# Innovation by Displacement

Linzhuo Li[a], Yiling Lin[b], and Lingfei Wu[b]

**Abstract**

New ideas are often thought to arise from recombining existing knowledge. Yet despite rapid publication growth—and expanding opportunities for recombination—scientific breakthroughs remain rare. This gap between productivity and progress challenges recombinant growth theory as the prevailing account of innovation. We argue that the limitation of this theory lies in treating ideas solely as complements, overlooking that breakthroughs often arise when ideas act as substitutes. To test this, we integrate scientist interviews, bibliometric validation, and machine learning analysis of 41 million papers (1965-2024). Interviews reveal that breakthroughs are marked not by novelty (Atypicality) alone but by their ability to displace dominant ideas (Disruption). Large-scale analysis confirms that novelty and disruption represent distinct innovation mechanisms: they are negatively correlated across domains, periods, team sizes, and paper versions. Novel papers extend dominant ideas across topics and attract immediate attention; disruptive papers displace them within the same topic and generate lasting influence. Hence, progress slows not from lack of effort but because most research extends rather than overturns ideas. Applying this perspective reveals distinct roles of theories and methods in scientific change: methods more often drive breakthroughs, whereas theories tend to be novel but rarely disruptive, reinforcing the dominance of established ideas.

[a]Department of Sociology, Zhejiang University
[b]School of Computing and Information, University of Pittsburgh

**Corresponding author(s):**
Linzhuo Li, Department of Sociology, Zhejiang University, Hangzhou, Zhejiang 310058, China. Email: newllqllz@gmail.com
Lingfei Wu, School of Computing and Information, University of Pittsburgh, Pittsburgh, PA 15260, USA. Email: liw105@pitt.edu

A central paradox of modern science is that, despite explosive growth in publications, progress appears to be slowing (Bloom et al., 2020; Milojević, 2015; Park et al., 2023). Moore's Law offers a vivid illustration: for decades, the number of transistors on a chip doubled roughly every two years, yet sustaining this pace now requires more than eighteen times as many researchers as in the early 1970s (Bloom et al., 2020). This paradox between productivity and progress raises a fundamental question: what drives scientific breakthroughs?

The prevailing answer is the recombinant growth theory, which holds that new ideas arise from novel combinations of existing knowledge. As knowledge accumulates, the space of possible combinations expands, fueling accelerating innovation (Weitzman, 1998). This logic underpins influential measures of novelty in patents (Fleming, 2001) and scientific papers (Uzzi et al., 2013). Yet the theory fails to explain why, despite expanded opportunities for recombination, breakthroughs remain rare. To reconcile this puzzle, scholars have proposed several explanations: the exhaustion of "low-hanging fruit" (Cowen, 2011; Gordon, 2017), rising specialization that narrows the search space (Jones, 2009), and funding practices that discourage risk-taking (Wang et al., 2017). Despite their differences, these accounts share the recombinant growth assumption, framing stagnation as a failure to search widely. This view, however, conflicts with evidence that modern information and communication technologies allow scientists to search both deeper in history (Larivière et al., 2007) and more broadly across domains (Agrawal et al., 2018).

Recent evidence points to a different explanation of the productivity–progress paradox. Chu and Evans (2021) describe "durable dominance" in large, mature fields, where canonical works persistently crowd out newer contributions. Bradford's (1976) protein quantification method—still among the most cited papers in molecular biology nearly fifty years later—illustrates this endurance, both for its utility and because it has never been displaced. At the level of individual papers, Park et al. (2023) document a similar pattern. Using the Disruption Index, which measures the extent to which a paper shifts citation attention away from its references (Funk & Owen-Smith, 2017), they show a steady decline in disruption over recent decades, signaling a systemic shift from displacing to consolidating dominant ideas as the prevailing mode of research.

This evidence motivates an alternative perspective: breakthroughs may arise not from recombination but from displacement. Accumulation without displacement produces a densifying citation network in which core knowledge remains entrenched even as the frontier expands (Cole, 1983). Papers can appear novel by applying established ideas in new contexts, yet never challenge the foundations they cite. The result is novelty at the frontier but stagnation at the core—a structural account of the productivity–progress paradox. From this perspective, productivity fails to yield progress not because scientists search too narrowly, but because they search in the wrong way—extending rather than overturning paradigms (Kuhn, 1962). In other words, the problem lies less in the scope of search than in its *orientation*.

These insights resonate with philosophical and sociological theories of scientific change. Kuhn (1962) distinguished "normal science," which extends prevailing paradigms, from "scientific revolutions" that displace them. Merton (1968) described "obliteration by incorporation," where foundational ideas become so deeply absorbed they cease to be cited. Collins (1998) portrayed science as a zero-sum competition for limited attention, governed by the "law of small numbers," in which only a handful of ideas, individuals, or programs achieve lasting prominence. Together, these perspectives highlight replacement—rather than mere accumulation—as central to scientific change. Yet despite its importance in the philosophy and sociology of science, this tradition has been marginalized in quantitative studies of science for lack of scalable indicators and datasets. We argue that it can now be empirically tested through advances in the Science of Science, which integrates large-scale data, novel bibliometric measures, and machine learning (Fortunato et al., 2018; Wang & Barabási, 2021).

These ideas also resonate with innovation studies in economics and management. Schumpeter (1942) first described economic growth as a process of "creative destruction," in which new firms and technologies replace old ones. Building on this foundation, Aghion and Howitt (1992) formalized the dynamics of innovation-driven growth through successive technological replacement, while Christensen (1997) extended the concept to organizational settings, showing how "disruptive innovations" displace established incumbents. Our framework brings this logic to the scientific domain: progress in science arises when new ideas supplant dominant predecessors. By linking the sociology of knowledge with the economics of innovation, we offer a unified empirical account of creative destruction in science (McMahan & McFarland, 2021).

In this paper, we draw on theories of scientific change and empirical tools to examine the underexplored process of idea displacement. We focus on three interrelated questions:

**1. What defines a scientific breakthrough?** Drawing on global expert interviews and landmark cases selected by *Nature* for its 150th anniversary, we validate novelty (Uzzi et al., 2013) and disruption (Funk & Owen-Smith, 2017) as predictors of breakthroughs and assess the conditions that explain their performance. This analysis clarifies the nature of innovation and sets the stage for identifying the structural conditions of breakthroughs at scale.

**2. How are recombination and displacement structurally distinct?** Using 41 million papers published between 1965 and 2024, we examine the correlation between novelty and disruption at scale. Leveraging machine learning on paper topics, we test whether novel and disruptive papers differ in how they engage dominant ideas—as captured by their most-cited references—and in the distinct influences they produce on subsequent research.

**3. How do theories and methods shape scientific change through these mechanisms?** We extend recent evidence that methods are often more disruptive than theories (Leahey et al., 2023). Using machine learning to classify papers as theories or methods, we analyze their novelty and disruption in groups and test whether these metrics account for their roles in sustaining the dominance of established ideas or overturning them.

By combining expert interviews, large-scale bibliometric analysis, and machine learning, we move beyond treating novelty and disruption as outcome indicators (Wu et al., 2022). Instead, we conceptualize them as distinct mechanisms of innovation: recombination extends dominant ideas across topics to generate novelty and attracts short-term attention, whereas displacement challenges dominant ideas within the same topic and produces long-term impact.

These dynamics help explain why productivity so often fails to yield progress—because scientists more frequently extend than overturn dominant paradigms—and account for the recent decline of disruptive research (Park et al., 2023). Applying this perspective to methods and theories reveals their distinct roles in scientific change: methods more often drive progress, whereas theories tend to be novel but rarely disruptive, reinforcing durable dominance (Chu & Evans, 2021). In doing so, our analysis provides new empirical leverage for classic sociological debates about how ideas emerge, gain legitimacy, and ultimately supplant those that came before.

**Two Views of Scientific Innovation: Recombination and Displacement**

Prior work often treats ideas as complements, assuming that broader search across diverse domains is more likely to yield surprising innovation (Weitzman, 1998). From this view, innovation is framed in terms of scope—broad search as innovative and local search as conservative (March, 1991). We argue that ideas can also be substitutes, in which case local search may be radically innovative. Local search often consolidates knowledge and thus remains conservative, but when it probes dominant ideas for failure points—anomalies they cannot explain—it can generate breakthroughs (Kaplan & Vakili, 2015; Kuhn, 1962). We therefore propose classifying innovation not by scope but by *orientation*.

*The Recombination View: Ideas as Complements*

The recombination view originates with Schumpeter (1934), who defined innovation as the creation of "new combinations" of existing resources. Schmookler (1966) emphasized cumulative, demand-driven innovation grounded in prior knowledge, and Griliches (1979) linked knowledge production to R&D spillovers. Together, these accounts share a core assumption: ideas are fundamentally complementary, and innovation arises by combining disparate but mutually reinforcing knowledge.

This logic was later formalized in organizational theory and economics. March (1991) distinguished between exploitation and exploration, arguing that only broad search across unfamiliar domains sustains long-term innovation. Building on this principle, Weitzman (1998) introduced a mathematical model of combinatorial growth, subsequently extended by Kauffman (2000) and Arthur (2009) to explain technological evolution through increasingly complex recombination.

Empirical studies reinforced this perspective. Fleming (2001) showed that broader recombination across invention categories predicted higher patent citations. Uzzi et al. (2013) extended this logic to science, finding that papers citing "atypical" journal pairings received more citations, especially when embedded within canonical literature. Other work emphasized variance rather than averages: Singh and Fleming (2010) and Wang et al. (2017) found that broad recombination produces greater variability in citation outcomes. These novelty measures remain indirect: citation counts capture applied or economic value, and citation variance reflects risk, but neither directly measure the production of new ideas.

Even direct attempts to link recombination to conceptual novelty yield mixed results. Taylor and Greve (2006) found that genre mixing generated novelty in comic books, but only when authors had prior collaborations. Leahey et al. (2017) showed interdisciplinary teams were most effective when built on repeated collaboration. Kaplan and Vakili (2015) found that breakthrough patents in nanotechnology were more likely to arise from local rather than broad search. Fontana et al. (2020) showed that common indicators of combinatorial novelty—such as atypicality (Uzzi et al. 2013)—track interdisciplinarity but poorly predict recognized breakthroughs like Nobel Prize papers.

Taken together, recombination offers a compelling metaphor and modeling tradition, but the empirical evidence is mixed. The view that breakthroughs emerge from novel, complementary ideas explains some cases—especially when supported by effective collaboration—but it leaves important gaps, underscoring that novelty alone is insufficient to generate breakthroughs.

*The Displacement View: Ideas as Substitutes*

In contrast to recombination's assumption of complementarity, the displacement view emphasizes substitution as the driving force of scientific change. Innovation here is understood not as the product of broad search, but as the result of focused efforts to challenge and ultimately replace dominant ideas.

Popper ([1934] 2002) emphasized falsification as a driver of progress, whereby new theories arise by disproving older ones. Fleck ([1935] 1981) described how thought styles persist until disrupted by conceptual shifts. Schumpeter's ([1942] 2013) notion of "creative destruction" framed innovation as the systematic displacement of outdated ideas. Planck (1950) famously observed that "*a new scientific truth does not triumph by convincing its opponents... but rather because its opponents eventually die, and a new generation grows up that is familiar with it,*" highlighting generational turnover as a social condition for change—a dynamic often summarized as "science advances one funeral at a time."

Kuhn (1962) popularized this displacement logic, arguing that science advances not through incremental accumulation but through paradigm shifts. He distinguished "normal science," which works within a stable framework, from "revolutionary science," when anomalies destabilize that framework and demand its replacement. Scientific revolutions, he argued, occur when alternative paradigms render prior theories obsolete—not through refinement, but through epistemic rupture.

Merton (1968) offered a complementary perspective with his concept of "obliteration by incorporation," in which foundational contributions disappear not by failure but by success—absorbed so deeply into common knowledge that they cease to be cited. Whereas Kuhn highlighted rupture, Merton emphasized assimilation as another pathway of disappearance.

Later theorists refined these dynamics. Schön ([1963] 2013) introduced "conceptual displacement," the reframing of existing ideas to address new problems, often through analogy. Mulkay (1974) highlighted how scientists borrow terms from other fields to generate new concepts, such as "computer memory" or "electromagnetic waves." Feyerabend (1970) championed epistemic pluralism, stressing the value of producing diverse and incommensurable ideas to expand the degrees of freedom for scientific change. Collins (1998) framed idea competition as a zero-sum game within limited intellectual attention spaces, emphasizing the "law of small numbers"—that only a handful of ideas, individuals, or programs can achieve lasting prominence—and arguing that intellectual succession is driven by strategic positioning within networks of legitimacy.

Yet despite its theoretical richness, displacement has been difficult to operationalize empirically. Large-scale bibliometrics emphasize knowledge relatedness and idea complementarity (Kessler, 1963; Small, 1973), but not substitution. Existing studies have attempted to measure idea substitution, but typically relying on analyzing special events and therefore cannot be scaled up. Chen (2004) introduced the concept of "pivot nodes" in citation networks—papers that redirect intellectual attention. Wu et al. (2016) examined attention-allocation mechanisms shaping idea turnover in online communities. Evans et al. (2016) developed a measure of "paradigmaticness" based on abstract text, showing that early-stage disciplines feature linguistic variance while mature ones converge on shared jargon. Azoulay et al. (2019) linked the deaths of elite scientists to shifts in citation networks, empirically confirming Planck's claim that "science advances one funeral at a time." Extending Schumpeter's logic of "creative destruction," McMahan & McFarland (2021) showed that scientific curation—such as review articles—can suppress citations to original work, further shaping patterns of idea replacement.

Together, these contributions suggest that epistemic change is better understood not as additive accumulation across domains, but as displacement within intellectual lineages. Yet existing tools for capturing this dynamic remain limited, highlighting the need for scalable metrics to identify when—and how—dominant ideas are supplanted.

## How Do Recombination and Displacement Drive Scientific Change?

Building on the literature reviewed above, we treat recombination and displacement as distinct mechanisms of innovation. We analyze breakthrough cases to develop general principles of scientific change, which in turn guide our hypotheses about how these mechanisms operate across contexts. As outlined in the introduction, our inquiry is organized around three questions: (1) What defines a scientific breakthrough? (2) How are recombination and displacement structurally distinct? (3) How do methods and theories shape scientific change through these mechanisms? Each question motivates one or more hypotheses, developed through theoretical reasoning and tested with bibliometric evidence.

### 1. What Defines a Scientific Breakthrough?

The recombinant growth theory views breakthroughs as products of ross-domain recombination (Uzzi et al., 2013; Weitzman, 1998). Yet many transformative advances arise from focused engagement with established ideas. Alan Turing's 1936 paper, *On Computable Numbers, with an Application to the Entscheidungsproblem*, exemplifies this dynamic. His insight—that computation is a universal process that any repeatable mechanism can execute without intelligence—laid the foundation for today's digital infrastructure.

This idea did not result from blending disparate literatures. A close reading of Turing's seven references shows a highly conventional pattern rooted in mainstream mathematics and logic. These references included: (1) Gödel's 1931 article introducing the incompleteness theorems; (2) and (3) Church's two 1936 papers published in *American Journal of Mathematics* and *Journal of Symbolic Logic*; (4) Kleene's 1935 article on recursive function theory; and three textbooks: (5) Hobson's book *Theory of Functions of a Real Variable* (1907); (6) Hilbert and Ackermann's book *Principles of Theoretical Logic* ([1931] 1999); and (7) Hilbert and Bernays' book *Foundations of Mathematics* ([1934] 1968). For a young logician, citing Gödel, Church, Hilbert, and Kleene was not just common—it was expected.

What made Turing's local search so generative was the way he challenged the prior ideas. In citing Gödel, Turing did not simply extend the framework of self-contained formal logics. Observing that formal logic lacked a precise definition of computational cost, he redefined what counted as logic itself—shifting its foundation from symbolic systems to mechanical computation. In doing so, he introduced an entirely new intellectual system based on machines and procedures rather than axioms and theorems.

This case underscores the limits of the recombination growth theory in explaining breakthroughs, because the surprise lies not in the scope of search but in its orientation—whether scholars extend existing paradigms or probes them for anomalies they cannot explain (Kaplan & Vakili 2015; Kuhn, 1962). By Uzzi et al.'s (2013) metric of journal reference novelty, Turing's paper would appear conventional. Yet it catalyzed one of the most consequential innovations of the twentieth century: the modern computer.

Building on this insight, we return to a central limitation of recombinant growth theory: its treatment of ideas solely as complements, overlooking cases where they act as substitutes. Turing's example illustrates this limitation in historical context and points to a more general principle we call *functional equivalence*: displacement occurs when a new idea performs the same epistemic function as a dominant predecessor but in a more compelling, scalable, or tractable way. For a breakthrough to occur, such functionally equivalent ideas must not only emerge but also overturn their predecessors and gain lasting impact. These considerations motivate the following expectations:

*Hypothesis 1*: Breakthroughs are associated with the displacement of dominant ideas.

## 2. How Are Recombination and Displacement Structurally Distinct?

We argue that, when it comes to the displacement of dominant ideas required for breakthroughs, recombination is unlikely to generate such displacement and may, in fact, reduce its likelihood. By drawing together distant, complementary pieces of knowledge, recombination makes it difficult for the resulting idea to perform the same epistemic function—let alone outperform and replace a single predecessor.

To illustrate, consider a hypothetical example: combining a microwave and a television might yield a "microwave TV"—a device that entertains while heating food. Such a hybrid product may be novel, but it is unlikely to replace either the microwave as a culinary tool or the television as a media device. Recombination is thus creative but not substitutive. In other words, it produces something different, but not necessarily something better. Displacement, by contrast, offers a superior functionally equivalent solution, enabling a new idea to replace a dominant one.

Despite extensive literatures on both perspectives, direct comparisons between recombination and displacement as mechanisms of innovation are rare. One reason is that they have historically been examined at different levels of analysis. Recombination has been well operationalized at the micro level, in studies of individual papers (Uzzi et al., 2013) and patents (Fleming, 2001). Displacement, by contrast, has largely been explored at the macro level, through accounts of paradigm shifts in science (Kuhn, 1962) and structural transformations in technology and markets (Christensen, 1997; Schumpeter, 1942). We next review emerging metrics for each mechanism, providing a basis for direct comparison at a common level of analysis.

The recombination view has focused on the paper level and produced several influential indicators. Atypicality measures how unexpectedly journals are co-cited within a paper (Uzzi et al., 2013). The Rao–Stirling Index captures how references span conceptually distant subject categories (Leahey et al., 2017). More recent approaches, such as neural embedding models, extend these discrete measures of combinatorial novelty into continuous ones within semantic space (Peng et al., 2021).

Metrics of displacement, by contrast, are rare. Most related measures focus on the formation of new ideas rather than on the displacement of old ones, such as intellectual turning‑point identification (Chen, 2004), the ambiguity score (McMahan & Evans, 2018), paradigmaticness (Evans et al., 2016), the Meme Score (Kuhn et al., 2014), and the ForeCite Score (King et al., 2020). The main exception is the Disruption Index (Funk & Owen-Smith 2017; Wu et al., 2019), which captures the extent to which a paper shifts citation attention away from its references and thus measures substitution between ideas.

Among these measures, atypicality and disruption are especially well suited for systematic comparison. Both operate at the micro level, focusing on individual papers, but they capture different dynamics: atypicality reflects *ex ante* novelty from recombination—what a paper draws upon—whereas disruption reflects *ex post* impact from displacement—how a paper redirects attention away from prior work. Because recombination and displacement represent fundamentally different mechanisms of epistemic change—one additive, the other substitutive—their effects are unlikely to coincide. Indeed, the greater the functional distance among the ideas being recombined, the less likely the resulting idea is to replace them all. This structural tension leads us to expect divergence between the two measures. For these reasons, we propose the following:

*Hypothesis 2a*: Novelty and Disruption Are Negatively Correlated.

We further argue that the distinction between novel and disruptive papers becomes sharper when examined through the structural conditions that give rise to each process, particularly in relation to the dominant ideas they build upon—their most-cited reference. Turing's 1936 paper provides an illustrative case: displacement tends to emerge within the same topic as the major idea it cites—in this case, Gödel's 1931 article. Such topic proximity increases the likelihood of functional equivalence and thus the potential for displacement. Recombination, by contrast, more often occurs across topics, applying dominant ideas from one domain to new contexts without directly challenging their core assumptions.

This contrast reflects two ecologies of science. Recombination motivates "*big science*", characterized by costly instruments, large teams, and collective knowledge growth (De Solla Price, 1963). In this context, novelty arises through active collaboration and broad search, while progress stagnates when scientists fail to engage in it. Team-based research can help mitigate the rising burden of knowledge faced by individual scientists (Jones, 2009), but it also pushes the frontier further beyond the reach of any single researcher (Hall et al. 2018; Jones 2021).

Displacement, by contrast, aligns more closely with "*little science*"—individual work or smaller teams that disrupt prior knowledge (Wu et al., 2019) and enable selective forgetting (Candia & Uzzi, 2021). Here, progress stalls not because scientists fail to collaborate and search widely, but because these efforts reinforce established ideas and crowding out new ones (Chu & Evans, 2021). In this sense, the scientific enterprise can become "too big to innovate," much like large corporations (Christensen, 1997). Only by displacing dominant ideas—and thereby rendering the complementary applications built around them obsolete—can science become "smaller" and innovative again.

These two ecologies may coexist and reflect the dual faces of scientific progress, consistent with Kuhn's (1962) theory. During periods of normal science, knowledge systems expand to the point that innovation slows, constrained by the accumulated burden of prior work—an era that may describe recent decades, characterized by the decline of disruptive research (Park et al., 2023) and the durable dominance of canonical literature (Chu & Evans, 2021). By contrast, during scientific revolutions, the burden of knowledge is lifted, allowing newcomers to reach the frontier more quickly—as in the golden age of quantum physics in the 1920s (Jones, 2021). Based on these considerations, we propose the following:

*Hypothesis 2b*: Novel papers extend dominant ideas across topics.

*Hypothesis 2c*: Disruptive papers replace dominant ideas within the same topic.

A final structural distinction between recombination and displacement lies in their impacts on future inquiries. Recombination produces novelty by linking ideas across distant domains. Such combinations often generate direct applications or economic value, attracting immediate attention (Kaplan & Vakili, 2015), especially when rooted in conventional literature (Uzzi et al., 2013). Displacement, by contrast, represents a more dramatic form of epistemic change that unfolds more slowly, as it faces resistance from committed users, complementary applications, and existing infrastructures (Rogers, 1962). Its influence therefore tends to emerge over the long run (Bornmann & Tekles, 2019), as seen with "sleeping beauty" ideas in science (Ke et al., 2015; van Raan, 2004) and the J-curve trajectory of general-purpose technologies (Brynjolfsson et al., 2021). For these reasons, we propose the following:

*Hypothesis 2d*: Disruptive papers generate longer-term impact than novel papers.

*3. How Do Methods and Theories Shape Scientific Change Through These Mechanisms?*

Although the distinction between methods and theories is foundational to knowledge production (Abbott, 2004), their epistemic roles in scientific change—especially that of methods—remain underexplored. Classic frameworks—from Kuhn's (1962) paradigm shifts to Collins's (1998) epistemic cultures—typically cast theoretical change as the central event in scientific progress. Kuhn, for instance, highlighted the shift from the Ptolemaic to the Copernican model as a defining revolution. In this view, methods such as the invention of the telescope only serve a secondary, enabling function, while the displacement of theories marks the transformation.

Recent work by Leahey et al. (2023) provides an empirical turning point. Analyzing 2,540 articles from *Citation Classic* essays (1977-1993), they used keyword-based classification to identify papers introducing new methods, theories, or results. They find that methodological contributions tend to be more disruptive than baseline, whereas theoretical contributions are less disruptive. If this pattern holds at scale, methods may play a more central role in breakthroughs than traditionally acknowledged. As Leahey et al. note, "*New methods likely disrupt because they are so portable, have broader-than-imagined applications, and, importantly, rarely evolve in the application process.*" This emphasis in the value of methods resonates with Collins' (1998) account of how methodological advances drive science to become high-consensus, rapid-discovery—as exemplified by the telescope in 17th-century astronomy.

We build on these insights by positing that the innovative potentials of methods and theories reflect their distinct epistemic roles. Theories, as integrative frameworks for sense-making (Abbott, 2004), tend to be more novel. They often extend, nuance, or hybridize existing knowledge rather than replace it outright—especially in the social sciences, where contributions frequently situate themselves within ongoing debates rather than resolving them. As a result, theoretical innovations often exhibit higher ambiguity and are cited alongside, rather than in place of, their predecessors (McMahan & Evans, 2018). Methods, by contrast, are generative tools for problem solving (Abbott, 2004). They are designed as functionally equivalent alternatives to existing approaches, making them well suited for targeted displacement. In Latour's (1987) terms, methods are more "black-boxed" and portable across contexts, allowing them to substitute for one another.

*Hypothesis 3*: Theories are more novel but less disruptive than methods.

Together, these hypotheses provide a framework for understanding scientific change as a process of epistemic competition and replacement. By integrating sociological theory with computational data, we distinguish between recombination and displacement as mechanisms of innovation and propose a scalable, theory-driven approach to explaining how new ideas challenge—and ultimately replace—old ones.

## Data and Methods

We integrate scientist interviews, bibliometric analysis, and large-scale machine learning on scientific texts to test our hypotheses. To investigate what defines a scientific breakthrough, we analyze ten expert-nominated breakthroughs from a global survey alongside nine landmark papers highlighted in *Nature*'s 150th anniversary. To evaluate how recombination and displacement differentially shape scientific change, we analyze 41 million articles published between 1965 and 2024 in OpenAlex, computing each paper's novelty (atypicality) and disruption to assess their relationship across domains, time periods, team size, and paper versions. Finally, we apply large language models to classify impactful papers by contribution type—method or theory. We then trace how these types map onto novelty and disruption to clarify their distinct epistemic roles in scientific change.

*Data Sources*

We draw on three complementary datasets:

**Breakthrough Papers Identified by Experts.** Our first dataset of breakthrough papers comes from expert interviews. We reanalyzed data from our 2019 survey (Wu et al. 2019), which asked scientists across disciplines to identify papers that either consolidate or disrupt scientific knowledge. The survey was approved by the appropriate Institutional Review Board and conducted through in-person, phone, and online interviews. Respondents were asked to nominate three to ten papers in each of two categories: *consolidating* ("developmental") papers, which extend existing knowledge, and *disruptive* papers, which represent a significant departure from prior work. To anchor these definitions, we provided examples such as the first laboratory observation of Bose–Einstein condensation (Davis et al., 1995) as consolidating and the "self-organized criticality" model of complex systems (Bak et al., 1987) as disruptive. The participants included 20 researchers from ten leading institutions across five countries, including the United States, China, Japan, France, and Germany. These researchers represent nine disciplines spanning from mathematics and physics to psychology and economics. In total, they nominated 190 papers. The average D-index of nominated disruptive papers was 0.21 (top 1% of all 41 million papers in our dataset), while nominated consolidating papers averaged –0.011 (bottom 13%). This yields an area under the curve (AUC) of 0.83, showing strong alignment between expert judgment and the D-index as a measure of breakthroughs. From this pool of papers, we focus on the ten most frequently nominated disruptive papers for detailed analysis.

To broaden representation, we also analyze ten landmark papers selected by *Nature* editors for the journal's 150th anniversary (https://www.nature.com/collections/fajcgfjdgh). Of the ten highlighted papers, nine were successfully matched to our dataset. This set, which overlaps with our survey (e.g., Watson & Crick, 1953), provides an independent validation of the D-index: all nine papers score highly on disruption, with an average D-index of 0.15—again placing them in the top 1% of the 41 million papers in our dataset.

**Large-Scale Bibliometric Datasets.** We use OpenAlex, which comprises 49 million peer-reviewed journal articles published between 1800 and 2024. To ensure data quality, we restrict analysis to 41 million articles published between 1965 and 2024 with complete metadata on references, citations, venues, and field classifications. This dataset underpins our core analyses of disruption (D-index), calculated directly from OpenAlex, and atypicality, derived from the SciSciNet dataset (Li et al., 2024) and linked back to our journal articles. Paper metadata allows us to examine how these metrics vary across domains, periods, and team sizes.

To rule out potential confounders definitively, we examine metric relationships within the same scholarly work by identifying 2,461 linked version pairs in OpenAlex. These pairs arise from legitimate, field-normative practices—not duplicate submissions—including conference papers later expanded into journal articles and substantial revisions during the peer-review and publication process. Such multi-formats are common in computer science, applied mathematics, and biomedical research. Across these pairs, versions are separated by an average of 2.5 years, undergo major changes in text and references, and accumulate citations independently; for example, Rao and Hazzledine's 1999 conference paper on dislocation modeling had 11 references, whereas the 2000 journal version included 47 references. To ensure that citation patterns are not confounded by shared citations across versions, we examined overlap in their citing papers and found that 97.7% of pairs share none, with a mean Jaccard similarity of 0.0037, indicating that the community treats these versions as distinct contributions. These linked versions therefore offer a quasi-experimental setting to test how within-paper changes in reference structure—especially atypicality—predict changes in disruptive impact.

We use the Microsoft Academic Graph (MAG) scientific taxonomy (Shen et al., 2018), incorporated into OpenAlex metadata, to classify paper domains and compute alternative measures of novelty. The MAG taxonomy comprises a six-level hierarchy: level zero includes 19 broad research fields (e.g., *Mathematics*, *Biology*, *Chemistry*); level one contains 292 subfields; and levels two through five encompass 543,454 unique keywords and phrases. Each paper is linked to one or more labels predicted by a machine learning model, with each label assigned a probability value between zero and one indicating prediction confidence. In our analysis, when a paper has multiple labels, we select the one with the highest confidence score. For domain classification, we use the level-zero labels and merge them into three broad categories: Science & Engineering, Social Sciences, and Arts & Humanities. For measuring knowledge span as an alternative indicator of novelty, we use the level-one subfield labels (see Robustness Check).

To evaluate topical alignment in recombination and displacement, we focus on high-impact papers and construct a dataset of topical similarity using large language models. We embed each paper's title and abstract to generate vector representations, then compute cosine similarity between these vectors as a proxy for topical alignment (see Robustness Check for examples). Finally, we link these similarity measures to our innovation metrics (A-index, D-index) to conduct the main analyses.

**Method vs. Theory Papers Classified by Machine Learning.** To investigate how theories and methods differ in their epistemic roles, we use large language models (LLMs) to classify papers. From our main dataset of 41 million papers published between 1965 and 2024, we select 136,831 high-impact papers with more than 500 citations. Each paper is then classified as theory, method, or finding using Leahey et al.'s (2023) three-category scheme and the following prompt: *"For the following paper with abstract {paper abstract}, which of the following best characterizes its contribution: (A) theory, (B) method, (C) finding. Only respond with option A, B, or C. No explanation."* The LLM classified 73,540 papers as findings and 63,291 as theories or methods. Our analysis focuses on the latter group: 24,014 (38%) theories and 39,277 (62%) methods. This distribution is consistent with Leahey et al.'s study of 2,540 *Citation Classics*, where roughly two-thirds of papers were methods and one-third were theories. We further find that the share of "findings" decreases as citation thresholds rise. Findings account for 58% of papers with ≥300 citations, 46% with ≥1,000 citations, and only 19% with ≥10,000 citations. This pattern suggests that the most highly cited contributions are typically foundational—methods or theories—rather than empirical findings, reinforcing our focus on their roles in driving scientific change.

*Measuring Novelty: Idea Complementarity Across Conceptual Distance*

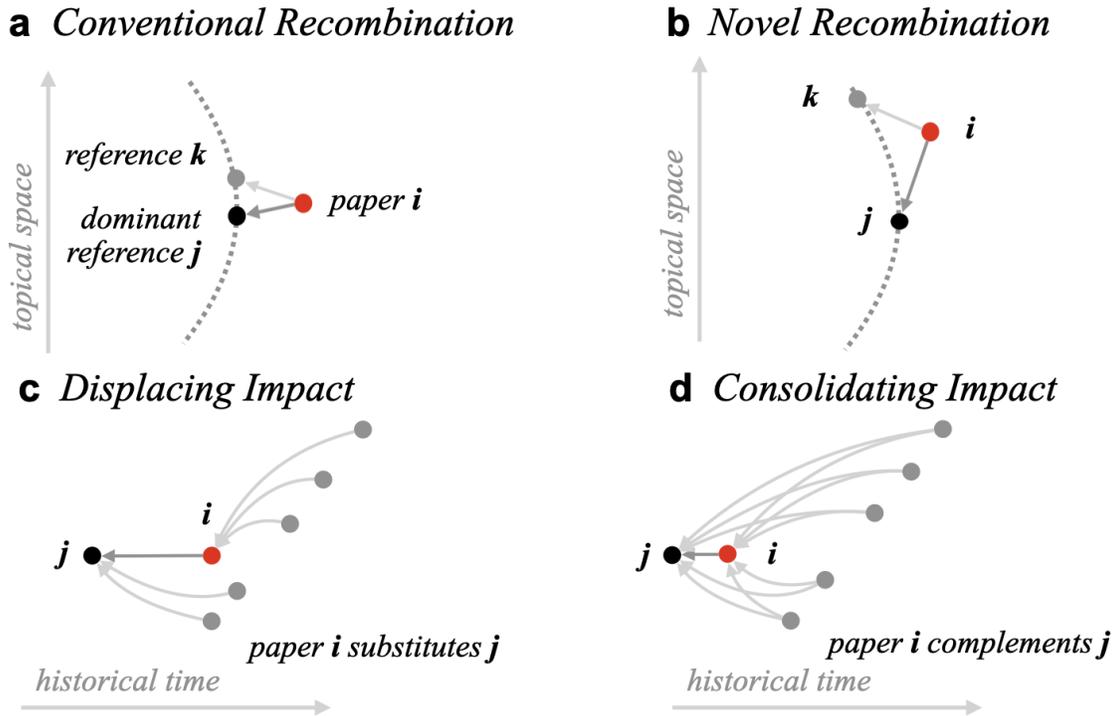

**Figure 1. Illustrations of Novelty and Disruption.** (a) *Conventional recombination*: Paper *i* (red) cites its dominant, most-cited reference *j* and other closely related works *k* located in the same dense region of the knowledge space, reinforcing established ideas. (b) *Novel recombination*: Paper *i* draws on references from more distant regions of the knowledge space, linking the dominant reference *j* with an unexpected source *k*, thereby extending established ideas into new contexts. (c) *Displacing impact*: In historical time, paper *i* substitutes its dominant reference *j*. Later works cite *i* but no longer cite *j*, indicating displacement of the prior dominant idea. (d) *Consolidating impact*: Paper *i* complements its dominant reference *j*. Later works cite *i* alongside *j*, reinforcing and extending the influence of established ideas rather than displacing them.

We adopt the atypicality introduced by Uzzi et al. (2013) to assess the combinatorial novelty of a paper. It captures the extent to which a paper cites unusual pairs of journals compared to random expectations. For each journal pair $(m, n)$, one can compute a z-score:

$$z_{mn} = \frac{obs_{mn} - exp_{mn}}{\sigma_{mn}}$$ Eq. (1)

where $obs_{mn}$ is the observed frequency of co-citation, $exp_{mn}$ is the expected frequency from randomized reference lists, and $\sigma_{mn}$ is its standard deviation. Higher $z_{mn}$ values indicate conventional pairings (e.g., *American Sociological Review* and *American Journal of Sociology*), while lower values reflect novel (e.g., *American Sociological Review* and *Physical Review Letters*). A paper's overall atypicality is measured as the 10th percentile of its pairwise z-scores, so novel papers have lower values than conventional papers, which is counterintuitive. For interpretability, we reverse the sign so that the Atypicality is higher for novel papers.

This z-score formulation is mathematically related to pointwise mutual information (PMI), a key concept in information science:

$$PMI_{mn} = log_2(\frac{P_{mn}}{P_m P_n}) = log_2(obs_{mn}) - log_2(exp_{mn})$$ Eq. (2)

where $P_m$ and $P_n$ represent the marginal probabilities of citing journals $m$ and $n$, and $P_{mn}$ is their joint probability. Both PMI and z-scores quantify how much more (or less) two items co-occur than expected by chance, capturing patterns of combinatorial novelty.

Recent advances in neural embedding models (e.g., word2vec, Mikolov et al., 2013; BERT, Devlin et al., 2018; Transformers, Vaswani et al., 2017) extend these discrete measures into continuous knowledge spaces. These models embed items—keywords, papers, or journals—into vector spaces where cosine similarity approximates PMI (Levy & Goldberg, 2014). In this light, atypicality can be seen as a measure of conceptual distance between complementary ideas across knowledge space (Miao et al., 2022; Peng et al., 2021; Tshitoyan et al., 2019). Our previous work further validates this equivalence between atypicality and conceptual distance using embedding models (Lin et al., 2022).

This insight into conceptual distance is useful for our research in two ways. First, because atypicality is derived from reference patterns, it may be partly shaped by citation practices that also influence disruption. This makes the observed relationship between novelty and disruption potentially confounded by citation practices, such as the increase in reference list length over time (Petersen et al., 2024). To address this concern, we construct an alternative measure of novelty using embeddings. We compute the cosine distance between embeddings of field labels assigned to a paper's references, yielding a measure of its *knowledge span*—the breadth of knowledge domains it integrates. This alternative measure allows us to test whether recombination across broader domains systematically undermines disruptive impact.

Second, this perspective allows us to move beyond journal-based atypicality to examine the conceptual distances among references generated through recombination, thereby testing our hypothesis that novel papers extend dominant ideas across topics.

*Measuring Disruption: Idea Substitution Over Historical Time*

We use the Disruption Index (D-index) to capture the displacement effect of a paper. Originally introduced by Funk and Owen-Smith (2017) for analyzing patent networks, this metric was adapted to scientific citations in our earlier work (Wu et al., 2019). Disruption measures the extent to which future work cites a focal paper *without* citing its references—signaling the eclipse of established knowledge.

For a focal paper, subsequent papers are categorized as: Type *a*: cites the focal paper but not its references (displacement); Type *b*: cites both the focal paper and its references (continuity); and Type *c*: cites only the references (no influence). The D-index of paper *p* is defined as

$$D_p = \frac{N_a - N_b}{N_a + N_b + N_c}$$
Eq. (3)

where $N_a$, $N_b$, and $N_c$ represent the counts of these three types of papers, respectively. Disruption values range from -1 (fully consolidating) to +1 (highly disruptive). Disruptive papers (D>0) attract citations that bypass their predecessors, while consolidating papers (D<0) are cited alongside them, reinforcing continuity.

Recent work shows that the D-index captures the displacement between a paper and its dominant (most-cited) reference, rather than its displacement relative to all references (Lin et al., 2025b). Indeed, as citations to the references follow a highly skewed, Zipfian distribution (Price, 1965; Wang et al., 2013), citations to non-dominant references are negligible. Therefore, the Disruption Index can be reformulated as:

$$D_p \approx d_p \frac{1}{1+b_p}$$
Eq. (4)

where $d_p = (N_a - N_b)/C_p$ is the local displacement factor, measuring whether a paper is cited independently ($N_a$) more often than alongside with its references ($N_b$); and $b_p = C_{max}/C_p$ is the *relative dominance factor*, capturing the citation impact of the dominant reference ($C_{max}$) relative to the focal paper ($C_p = N_i + N_j$). From this perspective, the D-index indicates whether a new idea successfully displaces the dominant idea it builds upon—its most cited reference (Figure 1 c-d). This insight motivates our empirical strategy: by analyzing the topical and temporal distance between disruptive papers and their most-cited references, we aim to better understand how dominant ideas are obsoleted and replaced within the evolving structure of knowledge.

### Analytical Strategies

We begin by analyzing expert-nominated papers to validate novelty (atypicality) and disruption as predictors of breakthroughs. We also evaluate whether these breakthroughs consolidate or displace their dominant references (*Hypothesis 1*). This analysis draws on interviews with twenty scientists across five countries and nine disciplines. To diversify selection criteria, we complement this set with landmark papers selected by *Nature* editors for its 150th anniversary.

We then turn to a large-scale analysis of 41 million journal articles published between 1965 and 2024. For each paper, we obtain its novelty and disruption. By examining their association, we test whether recombination and displacement represent distinct pathways to innovation (*Hypothesis 2a*). To ensure robustness, we stratify results by domain, time period, team size, and paper version. We further validate our results with an alternative novelty measure—knowledge span—generated by embedding models.

To assess how novel and disruptive papers engage with dominant ideas, we compare each paper to its most-cited reference in topical space. Using embedding models, we compute topic similarity between paper–reference pairs. For novel papers, we regress this similarity on the focal paper's novelty, controlling for potential confounders. We also test whether higher novelty reduces the similarity between a paper's most-cited reference and its other references. If novel papers extend dominant ideas across topics (*Hypothesis 2b*), we expect two patterns: first, as novelty increases, the focal paper becomes less similar to its most-cited reference, reflecting its movement into new topical space; second, because the remaining references constitute that new space, their similarity to the most-cited reference should also decrease. For disruptive papers, we conduct parallel analyses by regressing topic similarity on the focal paper's disruption. This tests whether disruptive papers tend to replace dominant ideas within the same topic (*Hypothesis 2c*). We do not examine the distance between the most-cited and remaining references here, since under displacement both are expected to remain close within the same topical space.

Next, we use the Sleeping Beauty Index (SBI) to measure long-term influence (Ke et al. 2015; van Raan 2004). This metric identifies papers that remain dormant for extended periods before receiving delayed recognition. We then compare SBI values between the most novel and most disruptive papers across disciplines, periods, and team sizes to test whether disruption generates longer-term impact than novelty (*Hypothesis 2d*).

Finally, we use large language models to classify papers as methods or theories and compare their average levels of novelty and disruption between these two groups (*Hypothesis 3*).

## Results

### Breakthroughs Are Associated With the Displacement of Dominant Ideas

Each breakthrough paper in our expert interview data cites a highly popular work as its dominant reference (Table 1), which is expected since widely cited papers are more likely to appear in reference lists (Cole & Cole, 1972). What sets these breakthrough papers apart is that each proposed solutions that ultimately displaced their predecessors.

Turing's (1936) paper exemplifies this dynamic. While Gödel's (1931) incompleteness theorems operated within symbolic logic, Turing mechanized the problem by proposing a model of computation that laid the foundation for computer science. Novelty (atypicality) misclassifies Turing's paper as conventional (A-index = –13.75), but disruption (D-index = 0.71) correctly captures its breakthrough nature.

Recent advances in AI follow the same pattern. Vaswani et al. (2017) introduced the Transformer architecture, which displaced Hochreiter and Schmidhuber's Long Short-Term Memory (LSTM) model (1997). While both are language models, Transformers scale more efficiently on modern hardware through parallelization, enabling dramatic gains in training speed and performance. As a result, despite LSTM accumulating 56,144 citations in our dataset, it has been eclipsed by Transformers, which has surpassed 80,575 citations and become the foundation of modern AI systems such as ChatGPT.

The meteoric rise of Transformers also sparked debates over credit. Schmidhuber argued that pioneers like him were "*denied adequate recognition for their contribution to the field of deep learning*" (Wikipedia contributors 2025), particularly after Hinton received the 2018 Turing Award and the 2024 Nobel Prize in Physics for neural network research. Such disputes are not unique: Duncan Watts and Steven Strogatz's (1998) "small-world" model drew similar critiques for eclipsing Milgram's (1967) original work. These cases illustrate a broader pattern in which displacement reshapes not only the epistemic foundations of a field (Cole, 1983) but also the allocation of recognition among scientists (Merton, 1968).

Another observation is that the pace of displacement varies: some unfold within months, while others take decades. Watson and Crick's (1953) double-helix model rapidly displaced Pauling and Corey's triple-helix structure within a year—an instance of intense competition during a high-consensus, rapid-discovery phase of science (Collins, 1998). By contrast, the Metropolis algorithm (1953), later formalized as Markov Chain Monte Carlo (MCMC), was only reframed into Simulated Annealing by Kirkpatrick et al. (1983) after three decades. Kirkpatrick and colleagues introduced a cooling schedule that improved optimization performance and extended the method's reach across engineering, machine learning, and operations research.

Across all ten cases, disruption consistently outperforms novelty in predicting breakthroughs. These breakthrough papers exhibit exceptionally high disruption scores (D-index = 0.58), far exceeding the 0.10 average observed for Nobel Prize–winning papers (Wu et al., 2019). Their extremely high citation counts—and placement in the very top citation percentiles within their fields—underscores disruption's value not only as a micro-level indicator but also as a marker of paradigm shifts at the field level. By contrast, novelty shows weaker predictive power: some breakthroughs are even misclassified as conservative (A-index < 0), highlighting that combinatorial novelty alone does not reliably predict transformative science (Fontana et al., 2020). Analysis of *Nature*'s collection of landmark papers yields consistent results (Table 3). Together, these findings support our first hypothesis: breakthroughs are marked not by novelty alone, but by their ability to displace dominant ideas (*Hypothesis 1*).

**Table 1. Breakthrough papers nominated in interviews and their most-cited references (shaded in gray).**

| Topic | Year | Authors | Paper Title | Citations | Disruption | Novelty (Atypicality) |
|-------|------|---------|-------------|-----------|------------|------------------------|
| DNA structure | 1953 | Watson & Crick | Molecular structure of nucleic acids: A Structure for Deoxyribose Nucleic Acid | Top 0.1% (6,311) | 0.96 | 0.71 |
|  | 1953 | Pauling & Corey | A proposed structure for the nucleic acids | 118 Top 1% |  |  |
| Fractal geometry | 1967 | Mandelbrot | How long is the coast of Britain? Statistical self-similarity and fractional dimension | Top 0.1% (2,419) | 0.95 | -4.68 |
|  | 1954 | Steinhaus | Length, shape and area | 88 Top 3% |  |  |
| Chaos theory | 1963 | Lorenz | Deterministic nonperiodic flow | Top 0.1% (12,213) | 0.81 | 2.27 |
|  | 1916 | Rayleigh | On convection currents in a horizontal layer of fluid | Top 0.1% (1,407) |  |  |
| Computation theory | 1936 | Turing | On computable numbers, with an application to the Entscheidungsproblem | Top 0.1% (2,700) | 0.71 | -13.75 |
|  | 1931 | Gödel | On formally undecidable theorems of Principia Mathematica and related systems I | Top 1% (229) |  |  |
| Network theory | 1998 | Watts & Strogatz | Collective dynamics of 'small-world' networks | Top 0.1% (23,626) | 0.53 | 46.09 |
| | 1967 | Milgram | The small world problem | Top 0.1% (4,934) | | |
| Neural networks | 2017 | Vaswani et al. | Attention Is All You Need | Top 0.1% (56,144) | 0.46 | NA[*] |
| | 1997 | Hochreiter & Schmidhuber | Long Short-Term Memory | Top 0.1% (80,575) | | |
| Text modeling | 2003 | Blei et al. | Latent Dirichlet allocation | Top 0.1% (10,340) | 0.42 | -24.37 |
| | 2000 | Nigam et al. | Text classification from labeled and unlabeled documents using EM | Top 0.1% (1,027) | | |
| Infant cognition | 1992 | Wynn | Addition and subtraction by human infants | Top 0.1% (815) | 0.36 | -3.12 |
| | 1980 | Starkey & Cooper | Perception of numbers by human infants | Top 1% (406) | | |
| Optimization algorithms | 1983 | Kirkpatrick et al. | Optimization by simulated annealing | Top 0.1% (24,395) | 0.29 | 17.36 |
| | 1953 | Metropolis et al. | Equation of state calculations by fast computing machines | Top 0.1% (24,156) | | |
| Game theory | 1951 | Nash | Non-cooperative games | Top 0.1% (3,491) | 0.28 | 0.05 |
| | 1944 | Von Neumann & Morgenstern | Theory of games and economic behavior | Top 0.1% (2,173) | | |

*Note: This paper primarily cites conference proceedings, whereas our atypicality index is based on journal co-citations. Its atypicality score is therefore biased, and we omit it here.

*Novelty and Disruption Are Negatively Correlated*

To evaluate whether the distinction between novelty and disruption generalizes beyond exemplary breakthroughs, we analyze bibliometric data covering 41 million papers published between 1965 and 2024. Across scientific domains, time periods, and team sizes, we find a robust negative correlation between the two measures (Figure 2). Disruption is especially common in papers from small teams (Wu et al., 2019) and earlier periods (Park et al., 2023), reinforcing the credibility of our results.

The most disruptive papers tend to draw on highly conventional references, often from core journals within their field. At first glance, this seems counterintuitive. In theory, one might expect disruptive work to arise from surprising journal pairings—for example, a sociology paper citing both *American Journal of Sociology* and *Physical Review Letters*—which broker across fields (Burt 2004). Later researchers might then struggle to trace the disparate sources and instead cite the focal paper alone, creating the appearance of disruption. Yet in practice, the opposite holds: highly disruptive work typically builds on conventional pairings such as *American Journal of Sociology* and *American Sociological Review*. This pattern suggests that the disruption index is not inflated by citing obscure or hard-to-trace references, but instead captures genuine innovation in the focal paper itself. Turing's (1936) paper—rooted entirely in mainstream mathematics and logic—illustrates this point. Taken together, these results indicate that breakthroughs often arise through deep, focused engagement with dominant ideas within a single intellectual tradition (Kaplan & Vakili, 2015).

By contrast, papers drawing on novel knowledge sources tend to show lower disruption—often falling below zero as novelty increases—indicating that novelty consolidates rather than displaces existing knowledge. This finding challenges the common assumption that breakthroughs primarily arise from bridging distant domains (Weitzman, 1998).

A key question is whether the negative correlation reflects a structural trade-off between novelty and disruption as distinct mechanisms, or simply differences in researcher style. Perhaps some scientists consistently pursue disruption by citing conventional sources, while others favor broader recombination but achieve less disruption. To address this concern, we examine 2,461 linked version pairs of the same scholarly work, separated by an average of 2.5 years. We find that as papers become more novel through revision, often by adding diverse references—a trend consistent with evidence that peer review encourages broader citation practices (Strang & Siler 2015). Correspondingly, their disruption declines (Figure 3).

Together, these results suggest that the negative correlation between novelty and disruption reflects a structural tension between two mechanisms of innovation rather than differences across scientists. Even when the same authors alter their citing practices, greater novelty reduces the likelihood that a paper will displace prior work, supporting *Hypothesis 2a*.

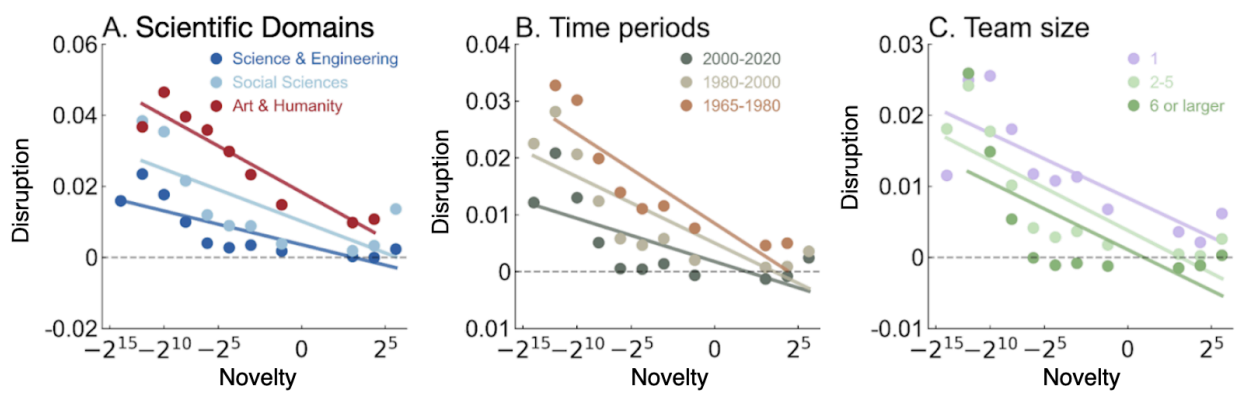

**Figure 2. Negative Correlation Between Paper Novelty and Disruption.** Based on an analysis of 40,935,251 research articles published between 1965 and 2020, we observe a consistent negative relationship between a paper's novelty (A-index) and its disruption (D-index). Each panel plots the average D-index across bins of A-index values, stratified by (a) scientific domains, (b) publication period, and (c) team size. In all cases, papers with highly atypical references (right end of the x-axis) tend to exhibit lower disruption scores. OLS regression lines are fitted to each category. All regression coefficients are negative and significant (p<0.01, two tailed Student's t-test)

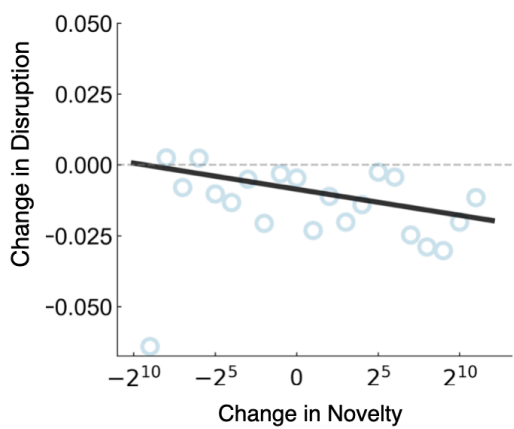

**Figure 3. Increasing a Paper's Atypicality Reduces Its Disruption.** We plot the change in disruption (D-index) against the change in reference novelty (A-index) across 2,461 papers with different versions, typically separated by 2.5 years and substantial revisions. Each point represents an aggregated bin of papers. The black line shows the OLS regression fit (coefficient = −0.0009, P < 0.05, two-sided Student's t-test).

*Novel Papers Extend Dominant Ideas Across Topics, Disruptive Papers Replace Dominant Ideas Within the Same Topic*

Using large language models, we compute cosine similarity between the embedding vectors of each focal paper and its most-cited reference—constructed from the title and abstract—to obtain a continuous measure of topical alignment (see Table 4 for examples). This measure allows us to assess how papers engage with dominant ideas within topical space.

We analyze 136,831 high-impact papers (≥500 citations) published between 1958 and 2020 from our dataset of 41 million papers. Topic similarity declines monotonically with novelty: papers in higher A-index deciles are about 13 percentile points less similar to their dominant references across the distribution. We also examine the relationship between novelty and the average topic similarity between a paper's most-cited reference and the centroid of its remaining references, and observe a parallel decline of about 16 percentile points (Figure 4).

Together, these patterns support *Hypothesis 2b*: novelty reflects the extension of dominant ideas across topics. They also suggest that novelty does not arise from indiscriminate mixing across all references, as implied by the "garbage can model" of innovation (Cohen et al., 1972). Instead, novelty reflects a systematic strategy of repurposing the dominant reference—for example, applying familiar ideas to unfamiliar contexts—consistent with Uzzi et al.'s (2013) theory of high-impact innovation.

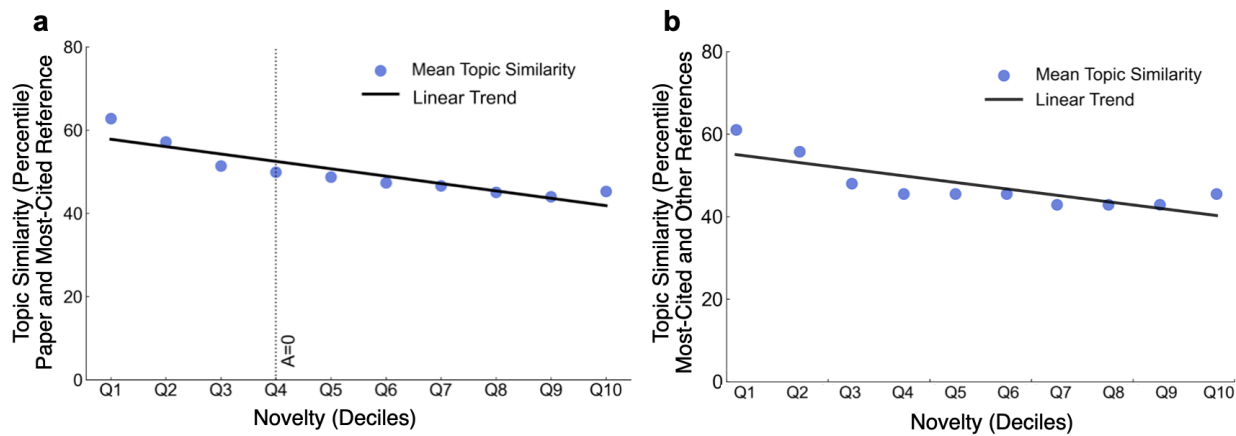

**Figure 4. Novel Papers Extend Dominant Ideas Across Topics.** We analyze 136,831 high-impact papers (≥500 citations) published between 1958 and 2020 in OpenAlex. Each paper's most-cited reference is identified as the "dominant idea," and its topic similarity to the focal paper and the centroid of its remaining references is computed using title and abstract embeddings (Qwen3-embedding model). (a) Topic similarity between the focal paper and the most-cited reference. Blue dots show decile-level means (Q1 = least novel, Q10 = most novel). The black line represents an OLS regression fit across all deciles (slope = −1.3, R² = 0.52, p < 0.05, two-sided t-test). (b) Topic similarity between most-cited references and remaining references. Blue dots show decile-level means, and the black line shows the regression fit (slope = −0.16, R² = 0.65, p < 0.01, two-sided t-test).

Using the same dataset, we next analyze the relationship between a focal paper's disruption and its topical similarity to the most-cited reference. The results reveal a striking V-shaped relationship (Figure 5). Among consolidating papers (D-index < 0), lower disruption scores correspond to higher topical similarity, consistent with the idea that consolidating work extends existing ideas within shared topical domains (Funk & Owen-Smith, 2017). A linear trend extrapolated from this region predicts continued decline in similarity as disruption increases. Yet among disruptive papers (D-index > 0), the trend reverses: higher disruption corresponds to greater similarity, indicating that disruption arises from displacing dominant ideas within the same topical space.

We test this association more rigorously by regressing topic similarity on disruption, with both variables normalized to percentiles for interpretability. Controls include (1) subfield overlap, to account for disciplinary boundaries (Milojević, 2015); (2) citation differences between the focal paper and its reference, to capture clustering around popular topics (Rzhetsky et al., 2015); and (3) publication year differences, to adjust for temporal proximity in responding to shocks or societal needs (Hill et al., 2025). We also include decade, field, and team-size fixed effects to absorb variation in citation norms, disciplinary practices, and collaboration scale (Larivière et al., 2008; Park et al., 2023; Wu et al., 2019). Because Figure 5 indicates a nonlinear, V-shaped pattern, regressions are restricted to disruptive papers (D > 0), where the reversal occurs. Regression results for consolidating papers (D < 0) are reported in Appendix Table B.

Regression models confirm that topic similarity increases significantly with disruption, even after controlling for potential confounders. In the baseline model, moving from D = 0 to maximum disruption corresponds to a 65-percentile increase in topic similarity. The effect remains stable across specifications; even in the fully controlled model, the same shift yields a 29-percentile increase (Table 2). These results demonstrate that the link between disruption and topical proximity is not only statistically robust but also substantively meaningful. These findings provide strong support for *Hypothesis 2c*.

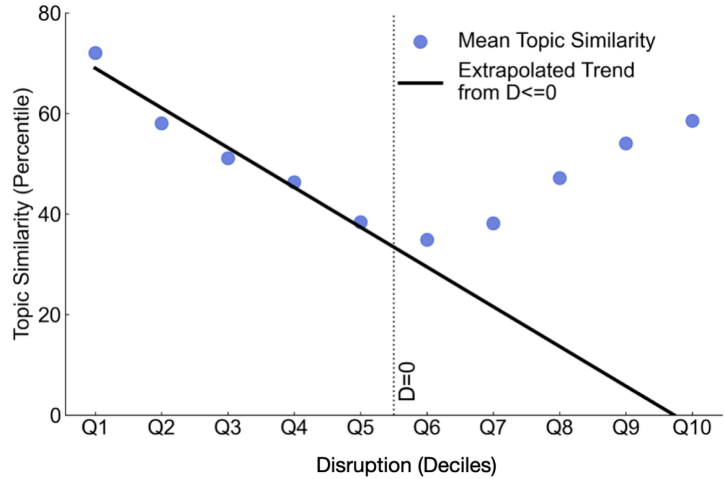

**Figure 5. Disruptive Papers Replace Dominant Ideas Within the Same Topic.** (a) Analyzing 136,831 journal articles (citations $\geq$ 500) published between 1958 and 2020, we observe a V-shaped pattern in topic similarity as disruption increases. Among consolidating papers (D < 0, Q1–Q5), higher D-index values are associated with lower topic similarity. The black line shows a linear fit across these five deciles (coefficient = −7.3, R² = 0.98, p < 0.01, two-sided t-test) and extrapolated to all deciles. Among disruptive papers (D > 0, Q6–Q10), the trend reverses: topic similarity rises with greater disruption. Blue dots represent mean cosine similarity scores between each focal paper and its most-cited reference, grouped by D-index decile from least (Q1) to most disruptive (Q10).

**Table 2. Regressions Predicting Topic Similarity from Disruption (D > 0).**

|  | Model 1 | Model 2 | Model 3 | Model 4 | Model 5 | Model 6 |
|---|---|---|---|---|---|---|
| D-index Percentile | 0.645*** | 0.642*** | 0.608*** | 0.603*** | 0.485*** | 0.292*** |
|  | (0.014) | (0.014) | (0.015) | (0.015) | (0.014) | (0.014) |
| Field Overlap |  |  |  |  | 0.189*** | 0.167*** |
|  |  |  |  |  | (0.003) | (0.003) |
| Citation Difference |  |  |  |  |  | -000*** |
|  |  |  |  |  |  | (0.000) |
| Year Difference |  |  |  |  |  | -0.003*** |
|  |  |  |  |  |  | (0.000) |
| Decade FE | No | Yes | Yes | Yes | Yes | Yes |
| Field FE | No | No | Yes | Yes | Yes | Yes |
| Team Size FE | No | No | No | Yes | Yes | Yes |
| Observations | 24,777 | 24,777 | 24,777 | 24,777 | 24,777 | 24,777 |
| R-squared | 0.081 | 0.081 | 0.109 | 0.111 | 0.214 | 0.263 |

Standard errors in parentheses. * p<0.05,** p<0.01 and *** p<0.001

*Disruptive papers generate longer-term impact than novel papers*

We next compare the long-run citation impact of novel and disruptive papers. From 41 million papers (1965–2020), we identify highly novel papers (top 5%, A-index > 23.4) and highly disruptive papers (top 5%, D-index > 0.02), yielding two comparable samples. To capture long-term impact, we use the Sleeping Beauty Index (SBI), which increases when papers remain dormant for extended periods before experiencing a surge of citation attention. Unlike raw citation counts, which reward immediate uptake, the Sleeping Beauty Index highlights works that shape their fields over the long run (Ke et al. 2015; van Raan 2004)

Figure 6 illustrates this dynamic by comparing annual citation trajectories for Watson and Crick's 1953 DNA paper with the average of 1,000 randomly selected papers. While breakthroughs like the paper by Watson and Crick sustain influence for decades, most papers peak quickly and fade, making "sleeping beauties" rare. Across scientific domains, publication periods, and team sizes, disruptive papers consistently exhibit higher SBI values than novel papers. Whereas novel papers generate short-lived bursts of attention, disruptive papers inspire subsequent research over much longer periods. This pattern supports *Hypothesis 2d*.

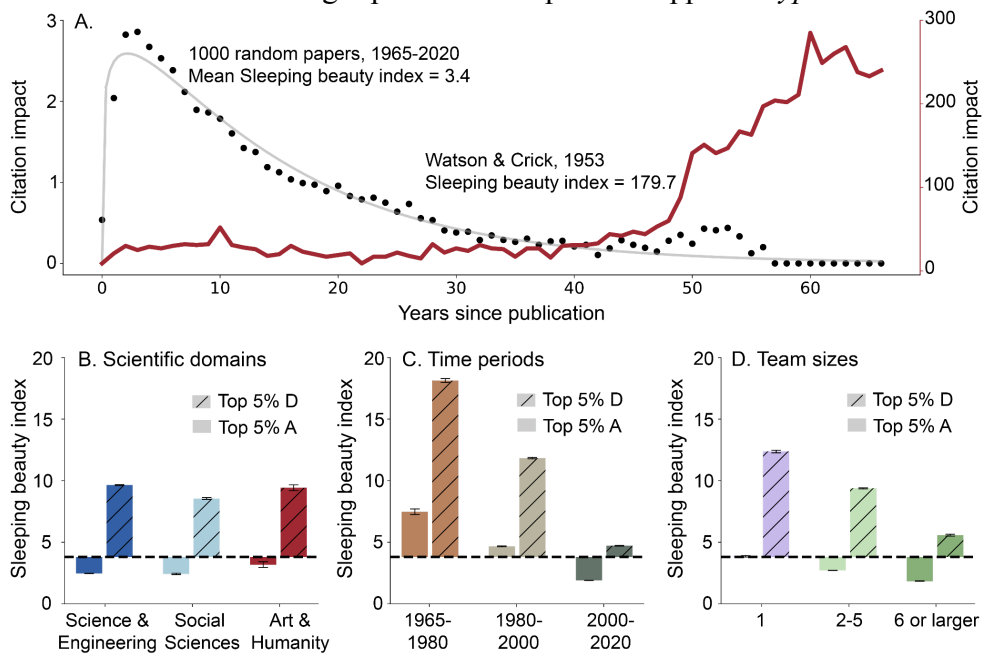

**Figure 6. Disruptive Papers Generate Greater Longer-Term Impacts Than Novel Papers.** Analysis of 41 million journal articles published between 1965 and 2020. (a) Annual citation trajectories for Watson and Crick's DNA paper compared with the average of 1,000 randomly selected papers, illustrating the rarity of long-term "sleeping beauties." (b–d) Comparisons between highly disruptive papers (top 5%, D-index > 0.02, hatched bars) and highly novel papers (top 5%, A-index > 23.4, solid bars) stratified by discipline, publication period, and team size. Across all contexts, disruptive papers consistently show higher SBI values, underscoring their greater long-term inspirational capacity. Horizontal lines mark the overall mean SBI (3.8) across all papers. Error bars denote 95% bootstrap confidence intervals.

When we examine the number of highly disruptive and novel papers over time, we find that the absolute number of highly disruptive papers has remained stable over the past six decades, despite rapid growth in overall publication volume—consistent with the "conservation of highly disruptive work" (Park et al., 2023). By contrast, the number of highly novel papers has increased dramatically in both absolute (Figure 7) and relative terms (Figure B1), showing that novel combinations are increasingly produced but rarely translate into breakthroughs. This pattern helps explain the productivity–progress paradox: scientists more often extend rather than overturn dominant paradigms, yielding diminishing returns to progress.

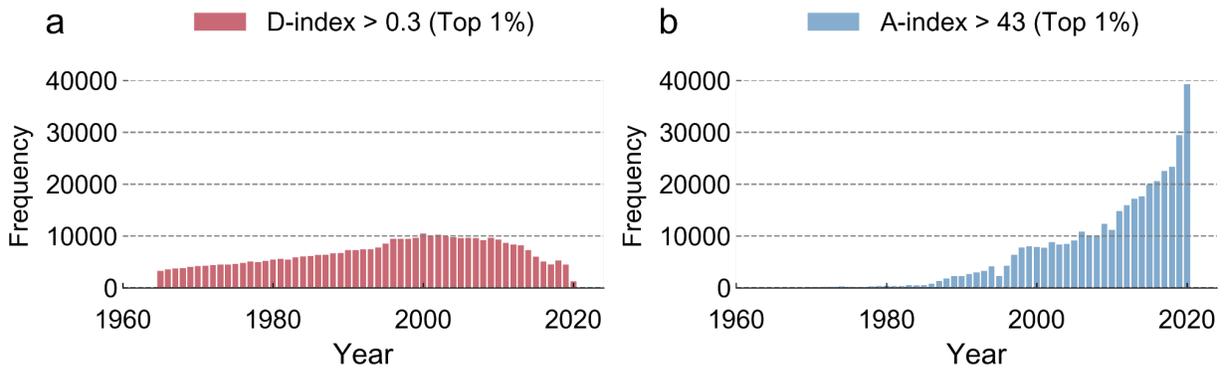

**Figure 7. Conservation of highly disruptive work.** From 1965 to 2020, the number of highly disruptive papers (top 1%, D-index > 0.3) remains stable, while the number of highly novel papers (top 1%, A-index > 43) grows rapidly.

*Theories are more novel but less disruptive than methods*

The clear distinction between novel and disruptive papers provides a foundation for analyzing the epistemic roles of theories and methods in scientific change. From our dataset of 41 million papers, we identified 63,291 high-impact papers (≥500 citations) that a large language model classified as either theories or methods. Linking these classifications to our computed metrics yielded 48,626 papers with available A-index and D-index scores. Of these, about one third are theories and two thirds are methods, consistent with Leahey et al.'s (2023) findings from a smaller sample (N=2,540). Theories exhibit higher novelty (A-index = –39.9) but lower disruption (D-index = 0.013), whereas methods exhibit lower novelty (A-index = –48.1) but higher disruption (D-index = 0.032).

To test these differences statistically, we regress novelty and disruption scores on a binary indicator of contribution type, controlling for field, decade, and team size. Results confirm that theoretical papers are significantly more novel than methodological ones (A-index difference = 8.17 units, ≈1 percentile, p < 0.01). In contrast, methodological papers are significantly more disruptive (D-index difference = 0.018 units, ≈5.1 percentiles, p < 0.001). These findings confirm *Hypothesis 3*: theories tend to be more novel, while methods tend to be more disruptive.

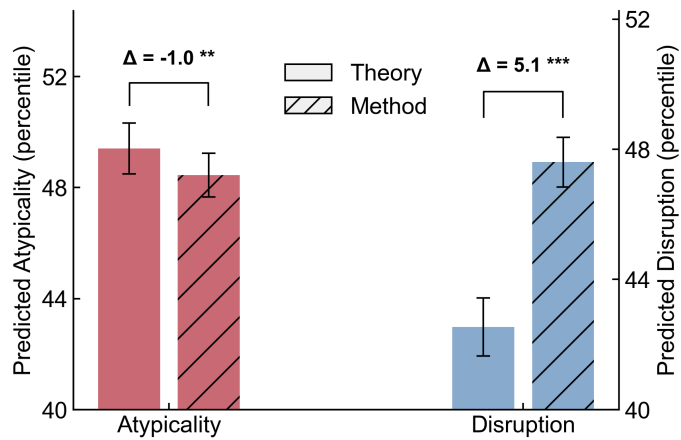

**Figure 8. Theoretical Papers Are More Novel but Less Disruptive Than Methodological Papers.** We identified 48,626 high impact papers (≥500 citations) as theory or method using a large language model. Compared to methodological papers (hatched), theoretical papers (solid) are significantly more novel (red) but less disruptive (blue). Predicted values are from OLS regressions with field, decade, and team size fixed effects. Error bars denote 95% bootstrap confidence intervals.

If theories tend to be novel but rarely disruptive, new theories will often reinforce rather than overturn established ideas, whereas new methods are more likely to displace them. Because theories typically engage with other theories and methods with other methods, this asymmetry should generate different dynamics: theoretical frameworks are more likely to achieve durable dominance, while methodological innovations are more likely to be replaced. We examine this empirically by analyzing the persistence of highly cited work across six decades (1965–2024). At each time point, we selected the 40,000 most-cited papers across all sciences, yielding 56,903 unique papers with abstracts in OpenAlex. We categorized these papers as theory (n = 11,564), method (n = 18,160), or findings (n = 27,184). For each time point, we then identified the top-k (here, k = 5,000) most-cited theories and methods separately and computed "dominance scores"—the Jaccard similarity between the top-k sets from consecutive decades. A score of 1 indicates complete persistence, while 0 indicates complete turnover.

Figure 9 shows that theory papers consistently exhibit higher persistence than method papers. This gap widened dramatically after 2015, reaching 17 percentage points by 2024 (theory = 0.63 vs. method = 0.46). The durable dominance of theories suggests that once established, they become deeply entrenched in scientific discourse. This reflects a form of intellectual path dependence: theories persist not only because they remain optimal, but because they are institutionally embedded in disciplinary epistemic culture, creating high switching costs (Cetina 1999; Collins, 1998). In contrast, methods are more vulnerable to replacement, as new techniques can more readily demonstrate superior performance or efficiency. In our analysis, this pattern holds robustly across different threshold values of *k* (100, 200, 500, 1000, and 2,000).

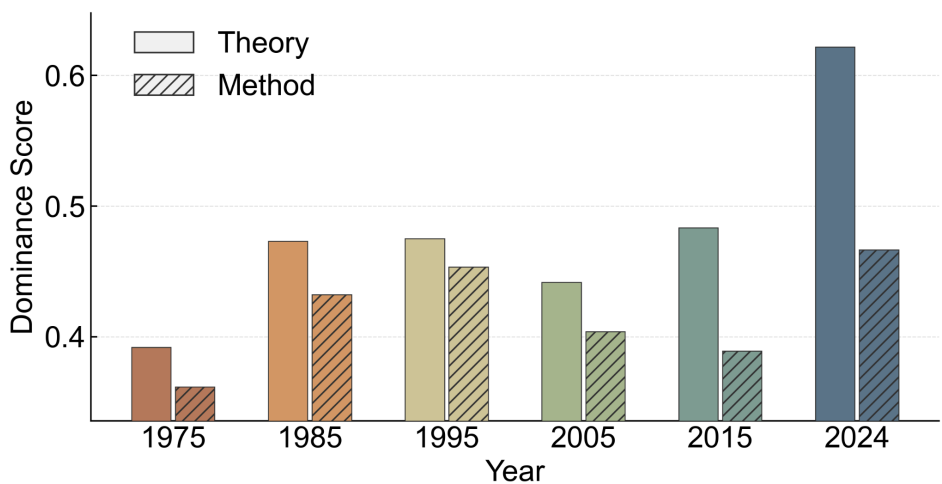

**Figure 9. Canonical theories persist longer than methods.** Analyzing 56,903 highly cited papers (1965–2024), we classified contributions as theory, method, or findings using large language models. We calculated "dominance scores" (Jaccard similarity of top-*k* most-cited papers across consecutive decades). Theory papers consistently show greater persistence than methods, with the gap widening after 2015 (2024: theory = 0.63 vs. method = 0.46). This suggests that theories endure longer in scientific discourse, while methods are more frequently displaced.

*Robustness Checks*

To ensure that the negative relationship between novelty and disruption is not an artifact of measurement or data limitations, we conduct several robustness checks.

**Expanded Case Analysis of Breakthrough Papers.**

Our main text analyzed ten expert-nominated discoveries to test whether breakthroughs arise through displacement rather than recombination. We extend this analysis by considering nine landmark studies highlighted by *Nature* editors for the journal's 150th anniversary (of the ten papers on this list, nine were matched to entries in OpenAlex). These cases span diverse domains but converge on the same result: disruption consistently outperforms novelty in identifying breakthroughs. Even when built on conventional reference recombination (A-index < 0), these studies show a clear orientation toward challenging dominant ideas (D-index > 0). For example, Köhler and Milstein's (1975) hybridoma technique (D = 0.69; A = –2.93) displaced earlier antibody assays; Farman et al.'s (1985) discovery of the ozone hole (D = 0.76; A = –3.27) overturned prevailing atmospheric models; and Kroto et al.'s (1985) identification of Buckminsterfullerene (D = 0.80; A = –9.49) redefined carbon structures. In each case, breakthroughs supplanted dominant paradigms with superior alternatives, redirecting citation attention away from their predecessors (Table 3).

**Table 3. Breakthrough papers selected by *Nature* editors for the journal's 150th anniversary and their most-cited references (shaded in gray).**

| Topic | Year | Authors | Paper Title | Citations | Disruption | Novelty (Atypicality) |
|---|---|---|---|---|---|---|
| DNA structure | 1953 | Watson & Crick | Molecular structure of nucleic acids: A Structure for Deoxyribose Nucleic Acid | Top 0.1% (6,311) | 0.96 | 0.71 |
| | 1953 | Pauling & Corey | A proposed structure for the nucleic acids | Top 1% (118) | | |
| Carbon molecule structure | 1985 | Kroto et al. | C60: Buckminsterfullerene | Top 0.1% (11,678) | 0.80 | -9.49 |
| | 1984 | Rohlfing et al. | Production and characterization of supersonic carbon cluster beams | Top 1% (681) | | |
| The ozone hole | 1985 | Farman et al. | Large losses of total ozone in Antarctica reveal seasonal ClOx/NOx interaction | Top 0.1% (2,227) | 0.76 | -3.27 |

| | 1978 | Dunkerton | On the mean meridional mass motions of the stratosphere and mesosphere | Top 1% (249 ) | | |
| Monoclonal antibodies | 1975 | Köhler & Milstein | Continuous cultures of fused cells secreting antibody of predefined specificity | Top 0.1% (10,178) | 0.69 | -2.93 |
| | 1963 | Jerne & Nordin | Plaque formation in agar by single antibody-producing cells | Top 0.1% (1,874) | | |
| Ion channels in cell membranes | 1976 | Neher & Sakmann | Single-channel currents recorded from membrane of denervated frog muscle fibers | Top 0.1% (1,115) | 0.37 | 0.39 |
| | 1973 | Anderson & Stevens | Voltage clamp analysis of acetylcholine produced end-plate current fluctuations at frog neuromuscular junction | Top 0.1% (735) | | |
| New subatomic particles | 1947 | Rochester & Butlers | Evidence for the Existence of New Unstable Elementary Particles | Top 1% (149 ) | 0.36 | -26.56 |
| | 1947 | Lattes et al. | Observations on the tracks of slow mesons in photographic emulsions | Top 1% (149) | | |
| Mesoporous materials | 1992 | Kresge et al. | Ordered mesoporous molecular sieves synthesized by a liquid-crystal template mechanism | Top 0.1% (13,241) | 0.21 | 54.26 |
| | 1967 | Gregg et al. | Adsorption surface area and porosity | Top 0.1% (8,222) | | |
| Planets outside the Solar System | 1995 | Mayor & Queloz | A Jupiter-mass companion to a solar-type star | Top 0.1% (2,628) | 0.21 | 3.51 |

| | 1991 | Duquennoy & Mayor | Multiplicity among solar-type stars in the solar neighbourhood | Top 3% (68) | | |
| Nuclear reprogramming | 1958 | Gurdon et.al | Sexually Mature Individuals of Xenopus laevis from the Transplantation of Single Somatic Nuclei | Top 1% (166) | 0.08 | -36.3 |
| | 1956 | King & Briggs | Serial transplantation of embryonic nuclei | Top 2% (91) | | |

**Alternative Measures of Novelty.** Our baseline measure of atypicality (Uzzi et al., 2013) evaluates novelty through journal pairings in reference lists. While influential, this metric is sensitive to citation norms. To test whether our findings generalize beyond citation-based definitions, we construct a content-based measure of novelty using semantic distances between referenced domains. Specifically, we draw level-one field-of-study labels from OpenAlex (292 fields, e.g., *Organic Chemistry*, *Discrete Mathematics*), embed them using large language models (SciBERT, Beltagy et al., 2019), and calculate each paper's "knowledge span" as the maximum pairwise distance between its referenced fields. Across 41 million papers (1965–2020), broader knowledge spans consistently predict lower disruption (Figure 10). This negative correlation holds across domains, periods, and team sizes, confirming the trade-off between recombination and displacement is robust to historical shifts in citation practices (Petersen et al., 2025).

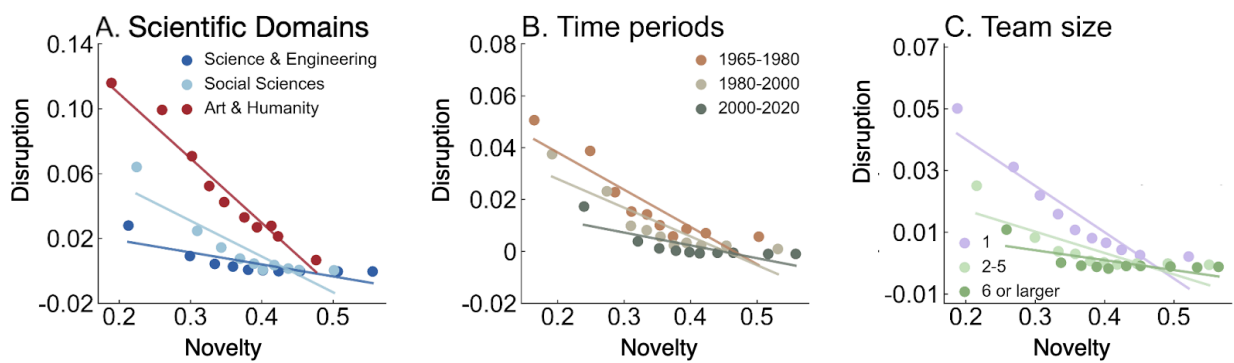

**Figure 10. Negative correlation between LLM-based (SciBERT) novelty and disruption.** Based on an analysis of 41 million research articles published between 1965 and 2020, we observe a consistent negative relationship between a paper's LLM-based atypicality (knowledge span) and its disruption (D-index). Each panel plots the average D-index across bins of knowledge span values, stratified by (A) scientific domains, (B) publication period, and (C) team size. Semantic distances are calculated using SciBERT embeddings of field-of-study labels assigned to each paper's references. In all cases, papers that recombine more distant domains (right end of x-axis) tend to exhibit lower disruption scores. All regression coefficients are negative and significant (p < 0.01, two-tailed Student's t-test).

To test model dependence, we replicate the analysis using two additional large language models with fundamentally different architectures, parameters, and training datasets: Google

Gemini (Figure 11) and GPT-2 (Figure 12). The distribution of knowledge-span values for each model is shown in Figure A1. Across all models, we observe the same negative relationship, confirming that the trade-off between novelty and disruption is robust to alternative LLM designs.

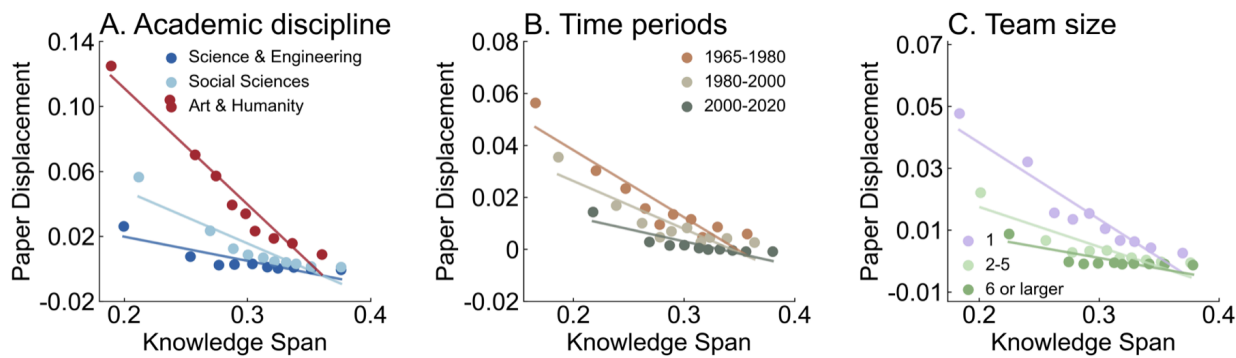

**Figure 11. Negative correlation between LLM-based (Gemini) novelty and disruption.** Analyzing 41 million journal articles (1965–2020), we find consistent negative correlations across domains, periods, and team sizes. A quasi-experimental analysis of 2,461 multi-version papers shows that increases in novelty correspond to declines in disruption, replicating the results in the main text.

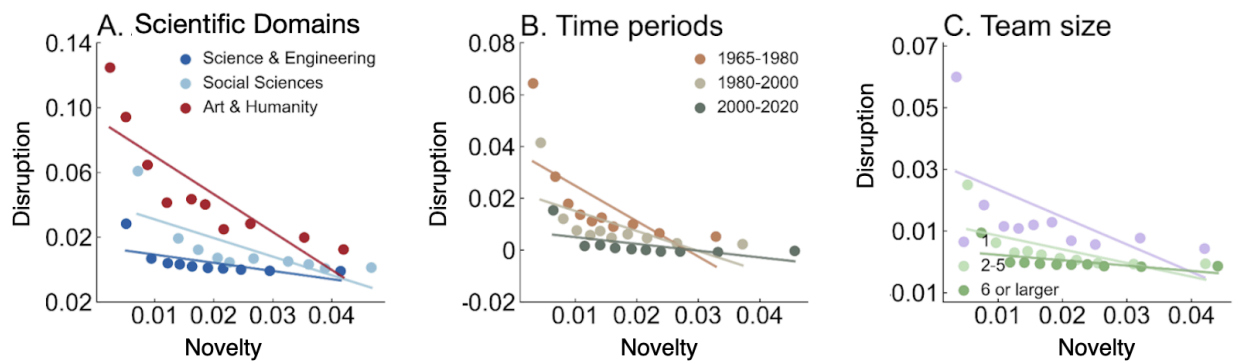

**Figure 12. Negative correlation between LLM-based (GPT-2) novelty and disruption.** Analyzing 41 million journal articles (1965–2020), we find consistent negative correlations across domains, periods, and team sizes. A quasi-experimental analysis of 2,461 multi-version papers shows that increases in novelty correspond to declines in disruption, replicating the results in the main text.

**Controlling for Citation Impact Levels.** A potential concern is that highly disruptive papers may also be highly cited, while highly novel papers may be less cited, making citation impact a confounder to the negative correlation between novelty and disruption. To test this, we stratify papers into three citation groups (1–10, 10–100, and 100+) and re-examine the relationship. In all groups, novelty and disruption remain negatively correlated (Figure 13). Thus, the observed trade-off is not driven by citation impact.

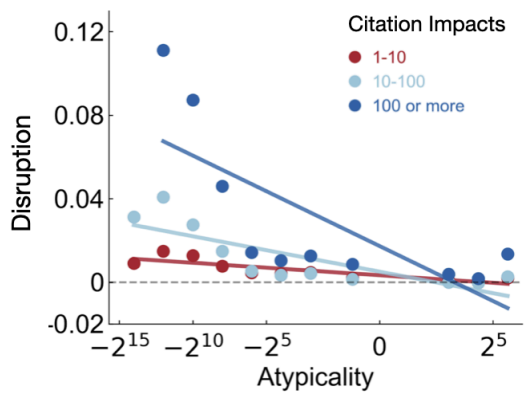

**Figure 13. Negative Correlation Between Paper Atypicality and Disruption Across Citation Impact Levels.** Based on an analysis of 40,935,251 research articles published between 1965 and 2020, we observe a consistent negative relationship between a paper's atypicality (A-index) and its disruption (D-index) after controlling for the citation impacts of the focal paper (1-10, 10-100 and 100 or more citations). OLS regression lines are fitted within each group. All regression coefficients are negative and significant (p<0.01, two tailed Student's t-test).

**Extended Analysis on Temporal Trends of Disruptive and Novel Papers.** We examine how the prevalence of novel versus disruptive papers has changed over time. Among 41 million papers (1965–2020), the share of disruptive papers (D-index > 0) declines from nearly 50% in 1965 to under 20% by 2020, while the share of novel papers (A-index > 0) rises from nearly 20% to over 50% (Figure 14). This divergence holds across domains, with the steepest decline of disruptive research in the humanities (Figure 15). These results complement the temporal trends in the absolute number of highly novel and disruptive papers presented in the main text, suggesting that while highly novel papers have proliferated in both absolute and relative terms, they rarely translate into breakthroughs.

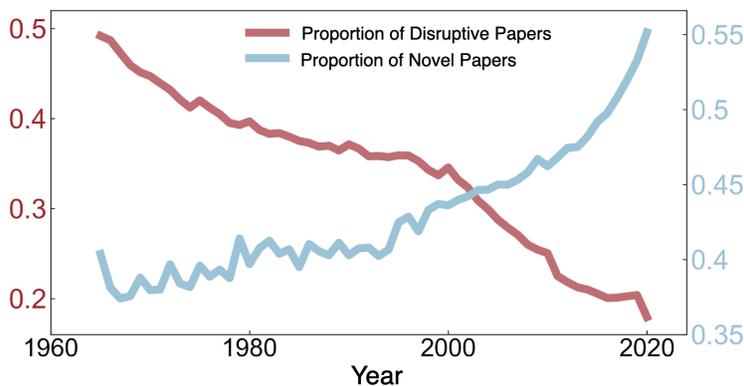

**Figure 14. Rise of novel papers and decline of disruptive papers.** Among 41 million journal articles (1965–2020), the share of displacing papers (D > 0) falls from ~50% to <20%, while the share of novel papers (A > 0) rises from ~20% to 50%.

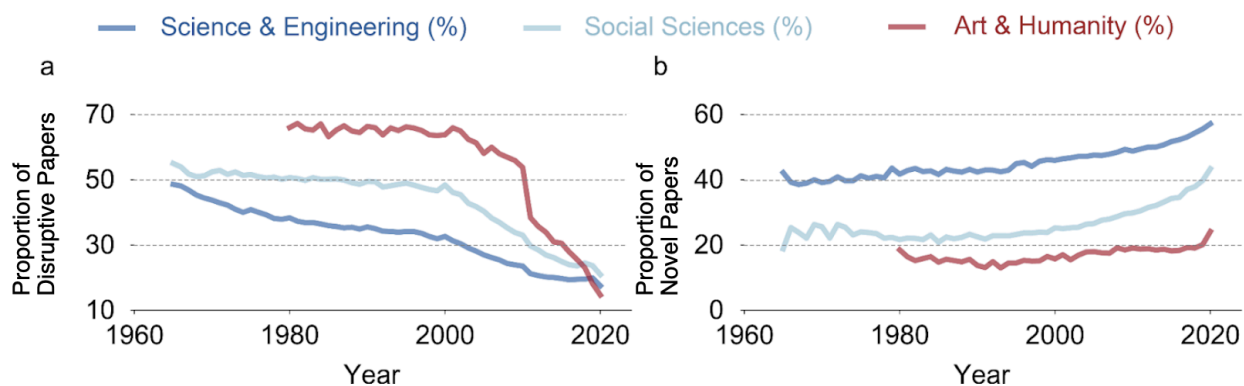

**Figure 15. Rise of atypical papers and decline of displacing papers across domains.** Trends hold in Science & Engineering, Social Sciences, and Arts & Humanities (1965–2020), with the steepest decline of disruptive research in Art & Humanities.

**Validation of LLM-based Topic Similarity.** To validate our LLM-based measure, we manually inspected focal papers and their most-cited references. A 1972 paper refining chromatographic methods for lichens and its 1970 predecessor address nearly identical problems, yielding high topical similarity (0.86) and strong disruption (0.31, placing it in the top 1% of all 41 million papers). By contrast, a 2019 ecology paper on bird migration citing Benjamini and Hochberg's (1995) statistics paper shows low similarity (0.24) and negligible disruption (0.000051). These examples confirm that the measure distinguishes close, domain-specific continuations from distant, cross-domain applications, supporting its validity for large-scale analysis.

**Table 4. Validation of LLM-based Topic Similarity.**

| Statistics | Paper | Paper Title and Abstract |
|---|---|---|
| Topic Similarity: 0.86<br><br>D-score: 0.31 | Focal paper (1972) | ***Improved conditions and new data for identification of lichen products by standardized thin layer chromatographic method.*** A standardized method for the identification of compounds by thin layer chromatography uses three solvent systems and RF classes coded on punched cards. This method, applied to the study of secondary products of lichen-forming fungi, has been improved and expanded to include easily prepared hydrolysis and O-methylation products. The data reported allow the confirmation of many substances previously difficult to identify and the proposal of structures for certain types of new compounds extracted from fragments of herbarium specimens. New data are given for 220 compounds and derivatives chromatographed in three standard solvent systems |
| | Most-cited Reference (1970) | ***A standardized method for the identification of lichen products.*** A procedure for the routine identification of the products of lichen-forming fungi by thin-layer chromatography is described. Microextracts of plant fragments are chromatographed in three solvent systems. The spots of unknowns are assigned to RF classes defined by the RF values of marker controls of two lichen substances (atranorin and norstictic acid) chromatographed on every plate. The unknowns are tentatively identified by sorting (by RF classes) punched cards summarizing microchemical data for all compounds previously studied. The preliminary identification is then confirmed by additional microchemical tests. The open-ended system can |
| | | incorporate new and unknown compounds as well as information from other chromatographic systems. Data obtained by the standardized procedure are given for 104 products. |
| Topic Similarity: 0.24<br><br>D-score: 0.000051 | Focal paper (2019) | ***Stochastic simulations reveal few green wave surfing populations among spring migrating herbivorous waterfowl.*** Tracking seasonally changing resources is regarded as a widespread proximate mechanism underpinning animal migration. Migrating herbivores, for example, are hypothesized to track seasonal foliage dynamics over large spatial scales. Previous investigations of this green wave hypothesis involved few species and limited geographical extent... Here, we introduce stochastic simulations to test this hypothesis using 222 individual spring migration episodes... We find that the green wave cannot be considered a ubiquitous driver of herbivorous waterfowl spring migration... highlighting key challenges in conserving migratory birds. |
| | Most-cited Reference (1995) | ***Controlling the false discovery rate: a practical and powerful approach to multiple testing.*** The common approach to the multiplicity problem calls for controlling the familywise error rate (FWER). This approach, though, has faults, and we point out a few. A different approach to problems of multiple significance testing is presented. It calls for controlling the expected proportion of falsely rejected hypotheses — the false discovery rate... A simple sequential Bonferroni-type procedure is proved to control the false discovery rate for independent test statistics... The use of the new procedure and the appropriateness of the criterion are illustrated with examples. |

**LLM Classification Robustness.** We test the robustness of our LLM-based classification of papers as theories or methods. Results are stable across three variations: (1) few-shot prompting with labeled exemplars (e.g., Turing/Gödel as theory, Watts/Milgram as method); (2) prompt rewording (substituting "theory/method" with "conceptual/formalism"); and (3) introducing a third "other" category to capture ambiguous cases. Across all strategies, classification patterns remain stable, confirming that LLMs can reliably distinguish epistemic roles at scale.

**Discussion and Conclusion**

Our findings challenge a prominent theory of scientific innovation—that breakthroughs arise mainly from broad combinations of disparate knowledge (Weitzman 1998). While recombination fosters creativity and novel perspectives, it overlooks a fundamental mechanism of innovation: the replacement of dominant ideas.

Across multiple empirical strategies, we document a persistent negative relationship between novelty and disruption. This pattern shows that novelty from recombination rarely leads to the displacement of entrenched paradigms. Instead, breakthroughs emerge when new work directly confronts—and ultimately supplants—the most central contributions of the past. Disruptive advances do not diversify their references; they focus narrowly on the core idea they aim to replace.

From this perspective, displacement offers a new explanation for the productivity–progress paradox: why breakthroughs remain rare despite rapid publication growth. Recombinant growth predicts accelerating progress through reconfiguration, yet our results suggest that recombination primarily fuels "normal science" rather than producing the paradigm shifts that mark "scientific revolutions" (Kuhn 1962). Breakthroughs are rare not because scientists lack effort, but because they more often extend than overturn existing ideas.

This insight carries direct implications for evaluation and funding. The dominant mode of scientific incentives favors incremental progress while discouraging field-redefining innovation (Alberts et al. 2014; Azoulay et al. 2011; Stephan 1996). Grant applicants are typically required to show how their work builds on prior contributions and promises broad impact (Azoulay & Li, 2021). Even the most ambitious reform proposals emphasize novelty or interdisciplinarity (e.g., Bromham et al., 2016; Wang et al., 2017), yet still operate within a recombinant framework. While "transformative science" is a stated goal (National Science Board 2007), few programs and mechanisms provide concrete pathways to achieve it or explicitly support the displacement of existing knowledge. We propose that addressing the productivity–progress paradox requires reorienting evaluation and funding systems toward paradigm-shifting ideas.

Similar concerns apply to AI. Systems such as AlphaFold (Jumper et al. 2021) or large language models like ChatGPT (Radford et al. 2019) excel at recombination (Agrawal et al. 2018). Trained on vast corpora, they generate outputs by detecting co-occurrence patterns and predicting likely continuations. This makes them powerful tools for extending knowledge but poorly suited for identifying, contesting, or replacing dominant paradigms. Without mechanisms for modeling displacement, AI risks reinforcing prevailing structures rather than catalyzing revolutions. Enabling AI to recognize and evaluate competing paradigms thus represents a critical frontier.

Our results also clarify earlier work. Lin, Evans, and Wu (2022) analyzed novelty, disruption, and citation impact using Uzzi et al.'s (2013) z-score framework, but blurred the distinction between novelty (10th percentile, A-index) and conventionality (median). By adopting the 10th-percentile measure, replicating and extending prior analyses, and incorporating an alternative LLM-based measure of novelty, we show a clear, robust negative correlation between novelty and disruption and demonstrate that they capture distinct innovation mechanisms.

In sum, science advances not only by building on past knowledge but also by selectively forgetting and replacing it. Displacement—alongside recombination—is a defining mechanism of breakthrough innovation. Recognizing this distinction reshapes how we understand progress, design indicators to measure it, and build institutions and technologies to support it.

## References

Abbott, Andrew. 2004. Methods of Discovery. New York, NY: WW Norton.

Aghion, P., & Howitt, P. (1992). A model of growth through creative destruction. *Econometrica, 60*(2), 323–351. https://doi.org/10.2307/2951599

Agrawal, Ajay, John McHale, and Alex Oettl. 2018. *Finding Needles in Haystacks: Artificial Intelligence and Recombinant Growth*. Cambridge, MA: National Bureau of Economic Research. doi: 10.3386/w24541.

Azoulay, Pierre, Christian Fons-Rosen, and Joshua S. Graff Zivin. 2019. "Does Science Advance One Funeral at a Time?" *The American Economic Review* 109(8):2889–2920.

Bak, P., C. Tang, and K. Wiesenfeld. 1987. "Self-Organized Criticality: An Explanation of the 1/f Noise." *Physical Review Letters* 59(4):381–84.

Bloom, Nicholas, Charles I. Jones, John Van Reenen, and Michael Webb. 2020. "Are Ideas Getting Harder to Find?" *American Economic Review* 110(4):1104–44.

Bornmann, Lutz, and Alexander Tekles. 2019. "Disruption Index Depends on Length of Citation Window." *El Profesional de La Información* 28(2). doi: 10.3145/epi.2019.mar.07.

Brian Arthur, W. 2009. *The Nature of Technology: What It Is and How It Evolves*. Simon and Schuster.

Burt, Ronald S. 2004. "Structural Holes and Good Ideas." *The American Journal of Sociology* 110(2):349–99.

Candia, Cristian, and Brian Uzzi. 2021. "Quantifying the Selective Forgetting and Integration of Ideas in Science and Technology." *The American Psychologist* 76(6):1067–87.

Chen, Chaomei. 2004. "Searching for Intellectual Turning Points: Progressive Knowledge Domain Visualization." *Proceedings of the National Academy of Sciences of the United States of America* 101 Suppl 1(suppl_1):5303–10.

Chu, Johan S. G., and James A. Evans. 2021. "Slowed Canonical Progress in Large Fields of Science." *Proceedings of the National Academy of Sciences of the United States of America* 118(41):e2021636118.

Cohen, Michael D., James G. March, and Johan P. Olsen. 1972. "A Garbage Can Model of Organizational Choice." *Administrative Science Quarterly* 17(1):1.

Cole, J. R., and S. Cole. 1972. "The Ortega Hypothesis: Citation Analysis Suggests That Only a Few Scientists Contribute to Scientific Progress." *Science (New York, N.Y.)* 178(4059):368–75.

Cole, Stephen. 1983. "The Hierarchy of the Sciences?" *The American Journal of Sociology* 89(1):111–39.

Collins, Randall. 1998. *The Sociology of Philosophies: A Global Theory of Intellectual Change*. London, England: Harvard University Press.

Cowen, T. 2011. *The Great Stagnation: How America Ate All the Low-Hanging Fruit of Modern History, Got Sick, and Will (eventually) Feel Better*. A Penguin eSpecial from Dutton.

Davis, K. B., M. Mewes, M. R. Andrews, van Druten NJ, D. S. Durfee, D. M. Kurn, and W. Ketterle. 1995. "Bose-Einstein Condensation in a Gas of Sodium Atoms." *Physical Review Letters* 75(22):3969–73.

De Solla Price, Derek J. 1963. *Little Science, Big Science*. Columbia University Press.

Devlin, Jacob, Ming-Wei Chang, Kenton Lee, and Kristina Toutanova. 2018. ""BERT: Pre-Training of Deep Bidirectional Transformers for Language Understanding.""

Dunkerton, Timothy. 1978. "On the Mean Meridional Mass Motions of the Stratosphere and Mesosphere." *Journal of the Atmospheric Sciences* 35(12):2325–33.

Evans, Eliza, Charles Gomez, and Daniel McFarland. 2016. "Measuring Paradigmaticness of Disciplines Using Text." *Sociological Science* 3:757–78.

Farman, J. C., B. G. Gardiner, and J. D. Shanklin. 1985. "Large Losses of Total Ozone in Antarctica Reveal Seasonal ClOx/NOx Interaction." *Nature* 315(6016):207–10.

Feyerabend, P. K. 1970. "Consolations for the Specialist." Pp. 197–230 in *Criticism and the Growth of Knowledge*, edited by I. Lakatos and A. Musgrave. Cambridge: Cambridge University Press.

Fleck, Ludwik. 1981. *Genesis and Development of a Scientific Fact*. Chicago, IL: University of Chicago Press.

Fleming, Lee. 2001. "Recombinant Uncertainty in Technological Search." *Management Science* 47(1):117–32.

Fontana, Magda, Martina Iori, Fabio Montobbio, and Roberta Sinatra. 2020. "New and Atypical Combinations: An Assessment of Novelty and Interdisciplinarity." *Research Policy* 49(7):104063.

Fortunato, Santo, Carl T. Bergstrom, Katy Börner, James A. Evans, Dirk Helbing, Staša Milojević, Alexander M. Petersen, Filippo Radicchi, Roberta Sinatra, Brian Uzzi, Alessandro Vespignani, Ludo Waltman, Dashun Wang, and Albert-László Barabási. 2018. "Science of Science." *Science (New York, N.Y.)* 359(6379). doi: 10.1126/science.aao0185.

Funk, Russell J., and Jason Owen-Smith. 2017. "A Dynamic Network Measure of Technological Change." *Management Science* 63(3):791–817.

Gödel, Kurt. 1931. "Über formal unentscheidbare Sätze der Principia Mathematica und verwandter Systeme I." *Monatshefte fur Mathematik* 38-38(1):173–98.

Gordon, Robert. 2017. *The Rise and Fall of American Growth*. Princeton: Princeton University Press.

Griliches, Zvi. 1979. "Issues in Assessing the Contribution of Research and Development to Productivity Growth." *The Bell Journal of Economics* 10(1):92.

Hicks, Diana, Paul Wouters, Ludo Waltman, Sarah de Rijcke, and Ismael Rafols. 2015. "Bibliometrics: The Leiden Manifesto for Research Metrics." *Nature* 520(7548):429–31.

Hill, Ryan, Yian Yin, Carolyn Stein, Xizhao Wang, Dashun Wang, and Benjamin F. Jones. 2025. "The Pivot Penalty in Research." *Nature* 642(8069):999–1006.

Jerne, N. K., and A. A. Nordin. 1963. "Plaque Formation in Agar by Single Antibody-Producing Cells." *Science (New York, N.Y.)* 140(3565):405.

Johnson, Steven. 2010. *Where Good Ideas Come from: The Natural History of Innovation.* Riverhead Books (Hardcover).

Jones, Benjamin F. 2009. "The Burden of Knowledge and the 'death of the Renaissance Man': Is Innovation Getting Harder?" *The Review of Economic Studies* 76(1):283–317.

Jones, Benjamin F. 2021. "The Rise of Research Teams: Benefits and Costs in Economics." *The Journal of Economic Perspectives: A Journal of the American Economic Association* 35(2):191–216.

Jumper, John, Richard Evans, Alexander Pritzel, Tim Green, Michael Figurnov, Olaf Ronneberger, Kathryn Tunyasuvunakool, Russ Bates, Augustin Žídek, Anna Potapenko, Alex Bridgland, Clemens Meyer, Simon A. A. Kohl, Andrew J. Ballard, Andrew Cowie, Bernardino Romera-Paredes, Stanislav Nikolov, Rishub Jain, Jonas Adler, Trevor Back, Stig Petersen, David Reiman, Ellen Clancy, Michal Zielinski, Martin Steinegger, Michalina Pacholska, Tamas Berghammer, Sebastian Bodenstein, David Silver, Oriol Vinyals, Andrew W. Senior, Koray Kavukcuoglu, Pushmeet Kohli, and Demis Hassabis. 2021. "Highly Accurate Protein Structure Prediction with AlphaFold." *Nature* 596(7873):583–89.

Kaplan, Sarah, and Keyvan Vakili. 2015. "The Double-Edged Sword of Recombination in Breakthrough Innovation." *Strategic Management Journal* 36(10):1435–57.

Kauffman, Stuart A. 2000. *Investigations.* Oxford University Press.

Ke, Qing, Emilio Ferrara, Filippo Radicchi, and Alessandro Flammini. 2015. "Defining and Identifying Sleeping Beauties in Science." *Proceedings of the National Academy of Sciences of the United States of America* 112(24):7426–31.

King, Daniel, Doug Downey, and Daniel S. Weld. 2020. "High-Precision Extraction of Emerging Concepts from Scientific Literature." in *Proceedings of the 43rd International ACM SIGIR Conference on Research and Development in Information Retrieval.* New York, NY, USA: ACM.

Kirkpatrick, S., C. D. Gelatt Jr, and M. P. Vecchi. 1983. "Optimization by Simulated Annealing." *Science (New York, N.Y.)* 220(4598):671–80.

Köhler, G., and C. Milstein. 1975. "Continuous Cultures of Fused Cells Secreting Antibody of Predefined Specificity." *Nature* 256(5517):495–97.

Kroto, H. W., J. R. Heath, S. C. O'Brien, R. F. Curl, and R. E. Smalley. 1985. "C60: Buckminsterfullerene." *Nature* 318(6042):162–63.

Kuhn, Thomas S. 1962. *The Structure of Scientific Revolutions: 50th Anniversary Edition.* University of Chicago Press.

Kuhn, Tobias, Matjaz Perc, and Dirk Helbing. 2014. ""Inheritance Patterns in Citation Networks Reveal Scientific Memes.""

Larivière, Vincent, Éric Archambault, and Yves Gingras. 2008. "Long-Term Variations in the Aging of Scientific Literature: From Exponential Growth to Steady-State Science (1900–2004)." *Journal of the American Society for Information Science and Technology* 59(2):288–96.

Latour, B. 1987. *Science in Action: How to Follow Scientists and Engineers through Society*. Harvard university press.

Leahey, Erin, Christine M. Beckman, and Taryn L. Stanko. 2017. "Prominent but Less Productive: The Impact of Interdisciplinarity on Scientists' Research." *Administrative Science Quarterly* 62(1):105–39.

Leahey, Erin, Jina Lee, and Russell J. Funk. 2023. "What Types of Novelty Are Most Disruptive?" *American Sociological Review* 88(3):562–97.

Levy, Omer, and Yoav Goldberg. 2014. "Neural Word Embedding as Implicit Matrix Factorization." Pp. 2177–85 in *Advances in neural information processing systems 27*, edited by Z. Ghahramani, M. Welling, C. Cortes, N. D. Lawrence, and K. Q. Weinberger. Curran Associates, Inc.

Lin, Yiling, James A. Evans, and Lingfei Wu. 2022. "New Directions in Science Emerge from Disconnection and Discord." *Journal of Informetrics* 16(1):101234.

Lin, Yiling, Linzhuo Li, and Lingfei Wu. 2025a. "Team Size and Its Negative Impact on the Disruption Index." *Journal of Informetrics* 19(3):101678.

Lin, Yiling, Linzhuo Li, and Lingfei Wu. 2025b. "The Disruption Index Measures Displacement between a Paper and Its Most Cited Reference." *Quantitative Science Studies*, forthcoming. arXiv:2504.04677.

Liu, Xin, Bu Yi, Ming Li, and Jiang Li. 2021. "Is Interdisciplinary Collaboration Research More Disruptive than Monodisciplinary Research?" *Proceedings of the Association for Information Science and Technology* 58(1):264–72.

March, James G. 1991. "Exploration and Exploitation in Organizational Learning." *Organization Science* 2(1):71–87.

McMahan, Peter, and James Evans. 2018. "Ambiguity and Engagement." *American Journal of Sociology* 124(3):860–912.

McMahan, Peter, and Daniel A. McFarland. 2021. "Creative Destruction: The Structural Consequences of Scientific Curation." *American Sociological Review* 86(2):341–76.

Merton, Robert K. 1968. *Social Theory and Social Structure*. 1968th ed. London, England: Macmillan.

Metropolis, Nicholas, Arianna W. Rosenbluth, Marshall N. Rosenbluth, Augusta H. Teller, and Edward Teller. 1953. "Equation of State Calculations by Fast Computing Machines." *The Journal of Chemical Physics* 21(6):1087–92.

Miao, Lili, Dakota Murray, Woo-Sung Jung, Vincent Larivière, Cassidy R. Sugimoto, and Yong-Yeol Ahn. 2022. "The Latent Structure of Global Scientific Development." *Nature Human Behaviour* 6(9):1206–17.

Mikolov, Tomas, Kai Chen, Greg Corrado, and Jeffrey Dean. 2013. ""Efficient Estimation of Word Representations in Vector Space.""

Milojević, Staša. 2015. "Quantifying the Cognitive Extent of Science." *Journal of Informetrics* 9(4):962–73.

Mulkay, Michael. 1974. "Conceptual Displacement and Migration in Science: A Prefatory Paper." *Science Studies* 4(3):205–34.

Park, Michael, Erin Leahey, and Russell J. Funk. 2023. "Papers and Patents Are Becoming Less Disruptive over Time." *Nature* 613(7942):138–44.

Pauling, L., and R. B. Corey. 1953. "A Proposed Structure for the Nucleic Acids." *Proceedings of the National Academy of Sciences of the United States of America* 39(2):84–97.

Peng, Hao, Qing Ke, Ceren Budak, Daniel M. Romero, and Yong-Yeol Ahn. 2021. "Neural Embeddings of Scholarly Periodicals Reveal Complex Disciplinary Organizations." *Science Advances* 7(17). doi: 10.1126/sciadv.abb9004.

Planck, Max K. 1950. *Scientific Autobiography and Other Papers*. New York, NY: Philosophical Library/Open Road.

Popper, Karl. 2002. *The Logic of Scientific Discovery*. 2nd ed. London, England: Routledge.

Porter, Theodore M. 1996. *Trust in Numbers: The Pursuit of Objectivity in Science and Public Life*. Princeton, NJ: Princeton University Press.

Price, Derek De Solla. 1976. "A General Theory of Bibliometric and Other Cumulative Advantage Processes." *Journal of the American Society for Information Science. American Society for Information Science* 27(5):292–306.

Radford, A., Wu, J., Child, R., Luan, D., Amodei, D., Sutskever, I. 2019. "Language Models Are Unsupervised Multitask Learners." *OpenAI Blog, 1(8), 9.*

Rohlfing, Eric A., D. M. Cox, and A. Kaldor. 1984. "Production and Characterization of Supersonic Carbon Cluster Beams." *The Journal of Chemical Physics* 81(7):3322–30.

Rzhetsky, Andrey, Jacob G. Foster, Ian T. Foster, and James A. Evans. 2015. "Choosing Experiments to Accelerate Collective Discovery." *Proceedings of the National Academy of Sciences of the United States of America* 112(47):14569–74.

Schmookler, J. 1966. *Invention and Economic Growth*. Cambridge, MA: Harvard University Press.

Schon, Donald A., ed. 2013. *Displacement of Concepts*. London, England: Routledge.

Schumpeter, Joseph A. 1934. *The Theory of Economic Development: An Inquiry into Profits, Capital, Credit, Interest, and the Business Cycle*. New Brunswick, New Jersey: Transaction Books.

Schumpeter, Joseph A. 2013. *Capitalism, Socialism and Democracy*. London, England: Routledge.

Singh, Jasjit, and Lee Fleming. 2010. "Lone Inventors as Sources of Breakthroughs: Myth or Reality?" *Management Science* 56(1):41–56.

Sinha, Arnab, Zhihong Shen, Yang Song, Hao Ma, Darrin Eide, Bo-June (paul) Hsu, and Kuansan Wang. 2015. "An Overview of Microsoft Academic Service (MAS) and Applications." in *Proceedings of the 24th International Conference on World Wide Web*. New York, NY, USA: ACM.

Strang, David, and Kyle Siler. 2015. "Revising as Reframing: Original Submissions versus Published Papers in *Administrative Science Quarterly*, 2005 to 2009." *Sociological Theory* 33(1):71–96.

Taylor, Alva, and Henrich R. Greve. 2006. "Superman or the Fantastic Four? Knowledge Combination And Experience in Innovative Teams." *Academy of Management Journal* 49(4):723–40.

Tshitoyan, Vahe, John Dagdelen, Leigh Weston, Alexander Dunn, Ziqin Rong, Olga Kononova, Kristin A. Persson, Gerbrand Ceder, and Anubhav Jain. 2019. "Unsupervised Word Embeddings Capture Latent Knowledge from Materials Science Literature." *Nature* 571(7763):95–98.

Turing, A. M. 1936. "On Computable Numbers, with an Application to the Entscheidungsproblem." *Proceedings of the London Mathematical Society. Third Series* s2-42(1):230–65.

Uzzi, Brian, Satyam Mukherjee, Michael Stringer, and Ben Jones. 2013. "Atypical Combinations and Scientific Impact." *Science* 342(6157):468–72.

Vaswani, Ashish, Noam Shazeer, Niki Parmar, Jakob Uszkoreit, Llion Jones, Aidan N. Gomez, Lukasz Kaiser, and Illia Polosukhin. 2017. ""Attention Is All You Need.""

Wang, Cheng-Jun, Lingfei Wu, Jiang Zhang, and Marco A. Janssen. 2016. "The Collective Direction of Attention Diffusion." *Scientific Reports* 6:34059.

Wang, Jian, Reinhilde Veugelers, and Paula Stephan. 2017. "Bias against Novelty in Science: A Cautionary Tale for Users of Bibliometric Indicators." *Research Policy* 46(8):1416–36.

Watson, J. D., and F. H. Crick. 1953. "Molecular Structure of Nucleic Acids; a Structure for Deoxyribose Nucleic Acid." *Nature* 171(4356):737–38.

Weitzman, M. L. 1998. "Recombinant Growth." *The Quarterly Journal of Economics* 113(2):331–60.

Wolfram, Stephen. 1984. "Universality and Complexity in Cellular Automata." *Physica D. Nonlinear Phenomena* 10(1-2):1–35.

Wuchty, Stefan, Benjamin F. Jones, and Brian Uzzi. 2007. "The Increasing Dominance of Teams in Production of Knowledge." *Science* 316(5827):1036–39.

Wu, Lingfei, Jacopo A. Baggio, and Marco A. Janssen. 2016. "The Role of Diverse Strategies in Sustainable Knowledge Production." *PloS One* 11(3):e0149151.

Wu, Lingfei, Aniket Kittur, Hyejin Youn, Staša Milojević, Erin Leahey, Stephen M. Fiore, and Yong-Yeol Ahn. 2022. "Metrics and Mechanisms: Measuring the Unmeasurable in the Science of Science." *Journal of Informetrics* 16(2):101290.

Wu, Lingfei, Dashun Wang, and James A. Evans. 2019. "Large Teams Develop and Small Teams Disrupt Science and Technology." *Nature* 566(7744):378–82.

Wu, Lingfei, and Jiang Zhang. 2013. "The Decentralized Flow Structure of Clickstreams on the Web." *The European Physical Journal. B* 86(6). doi: 10.1140/epjb/e2013-40132-2.

APPENDIX

*Appendix A. Measuring Knowledge Span using Large Language Models.*

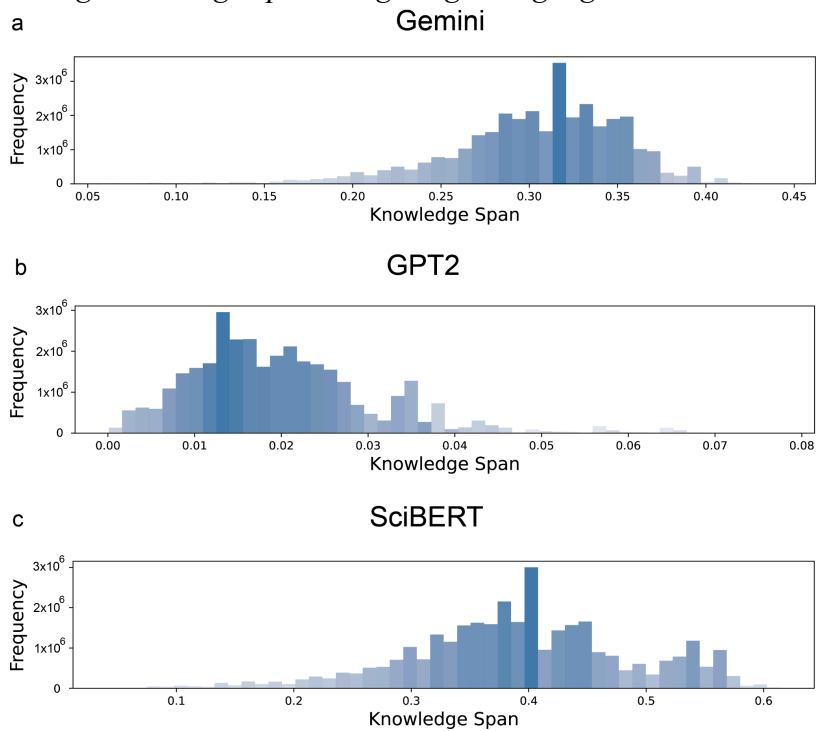

**Figure A1. Distribution of LLM-based conceptual novelty.** We measure the knowledge span of 41 million journal articles published between 1965 and 2020 using three different LLMs: (a) Google Gemini (embedding dimension = 768), (b) GPT-2 (dimension = 1024), and (c) SciBERT (dimension = 768). The knowledge span is defined as the maximum cosine distance among field-of-study embeddings in a paper's reference list. Although the distributions vary in scale, all three models capture the diversity of conceptual recombination across the corpus.

*Appendix B. Regressions Predicting Topic Similarity from Disruption (D-index < 0).*

| | Model 1 | Model 2 | Model 3 | Model 4 | Model 5 | Model 6 |
|---|---|---|---|---|---|---|
| D-index Percentile | -0.702*** | -0.696*** | -0.658*** | -0.658*** | -0.496*** | -0.343*** |
| | (0.009) | (0.009) | (0.009) | (0.009) | (0.009) | (0.01) |
| Field Overlap(True) | | | | | 0.223*** | 0.191*** |
| | | | | | (0.003) | (0.003) |
| Citation Difference | | | | | | -0.000*** |
| | | | | | | (0.000) |
| Year Difference | | | | | | -0.003*** |
| | | | | | | (0.000) |
| Decade FE | No | Yes | Yes | Yes | Yes | Yes |
| Field FE | No | No | Yes | Yes | Yes | Yes |
| Team Size FE | No | No | No | Yes | Yes | Yes |
| Observations | 32977 | 32977 | 32977 | 32977 | 32977 | 32977 |
| R-squared | 0.157 | 0.164 | 0.191 | 0.194 | 0.319 | 0.361 |

*Note*: Standard errors in parentheses. * p<.05,** p<.01 and *** p<.001 (two-tailed tests).